\documentclass[aps,prl,twocolumn,nofootinbib,longbibliography,amsfonts,amssymb,amsmath,superscriptaddress]{revtex4-2}
\usepackage[dvipsnames,table]{xcolor}
\usepackage{microtype}
\usepackage{IEEEtrantools}
\usepackage{mathrsfs}
\usepackage{amsthm}
\usepackage{enumitem}
\usepackage{varwidth}
\usepackage[normalem]{ulem}
\usepackage{graphicx}
\usepackage{stackengine}
\usepackage{cprotect}
\usepackage[caption=false]{subfig}
\usepackage{multirow}
\usepackage[pdftex]{hyperref}
\usepackage[USenglish]{babel}
\usepackage{microtype}

\DeclareMathOperator*{\argmax}{argmax}
\renewcommand{\sec}[1]{\emph{#1.---}}

\begin{document}

\title{Neural Importance Sampling for Rapid and Reliable Gravitational-Wave Inference}

\author{Maximilian Dax}
\thanks{Equal contribution}
\email{maximilian.dax@tuebingen.mpg.de}
\affiliation{Max Planck Institute for Intelligent Systems, Max-Planck-Ring 4, 72076 T\"ubingen, Germany}
\author{Stephen R. Green}
\thanks{Equal contribution}
\email{stephen.green@aei.mpg.de}
\affiliation{Max Planck Institute for Gravitational Physics (Albert Einstein Institute), Am M\"uhlenberg 1, 14476 Potsdam, Germany}
\author{Jonathan Gair}
\affiliation{Max Planck Institute for Gravitational Physics (Albert Einstein Institute), Am M\"uhlenberg 1, 14476 Potsdam, Germany}
\author{Michael P\"urrer}
\affiliation{Max Planck Institute for Gravitational Physics (Albert Einstein Institute), Am M\"uhlenberg 1, 14476 Potsdam, Germany}
\affiliation{Department of Physics, East Hall, University of Rhode Island, Kingston, RI 02881, USA}
\affiliation{URI Research Computing, Tyler Hall, University of Rhode Island, Kingston, RI 02881, USA}
\author{Jonas~Wildberger}
\affiliation{Max Planck Institute for Intelligent Systems,  Max-Planck-Ring 4, 72076 T\"ubingen, Germany}
\author{Jakob H.~Macke}
\affiliation{Max Planck Institute for Intelligent Systems,  Max-Planck-Ring 4, 72076 T\"ubingen, Germany}
\affiliation{Machine Learning in Science, University of T\"ubingen, 72076 T\"ubingen, Germany}
\author{Alessandra Buonanno}
\affiliation{Max Planck Institute for Gravitational Physics (Albert Einstein Institute), Am M\"uhlenberg 1, 14476 Potsdam, Germany}
\affiliation{Department of Physics, University of Maryland, College Park, MD 20742, USA}
\author{Bernhard Sch\"{o}lkopf}
\affiliation{Max Planck Institute for Intelligent Systems,  Max-Planck-Ring 4, 72076  T\"ubingen, Germany}

\begin{abstract}
  We combine amortized neural posterior estimation with importance sampling for fast and accurate gravitational-wave inference. We first generate a rapid proposal for the Bayesian posterior  using neural networks, and then attach importance weights based on the underlying likelihood and prior. This provides (1) a corrected posterior free from network inaccuracies, (2) a performance diagnostic (the sample efficiency) for assessing the proposal and identifying failure cases, and (3) an unbiased estimate of the Bayesian evidence. By establishing this independent verification and correction mechanism we address some of the most frequent criticisms against deep learning for scientific inference.  We carry out a large study analyzing 42 binary black hole mergers observed by LIGO and Virgo with the SEOBNRv4PHM and IMRPhenomXPHM waveform models. This shows a median sample efficiency of $\approx 10\%$ (two orders-of-magnitude better than standard samplers) as well as a ten-fold reduction in the statistical uncertainty in the log evidence. Given these advantages, we expect a significant impact on gravitational-wave inference, and for this approach to serve as a paradigm for harnessing deep learning methods in scientific applications.
\end{abstract}

\maketitle

\sec{Introduction}Bayesian inference is a key paradigm for
scientific discovery. In the context of gravitational waves (GWs), it
underlies analyses including individual-event parameter estimation~\cite{LIGOScientific:2021djp}, tests of
gravity~\cite{LIGOScientific:2021sio}, neutron-star
physics~\cite{Abbott:2018exr},
populations~\cite{LIGOScientific:2021psn}, and
cosmology~\cite{LIGOScientific:2021aug}. Given a prior $p(\theta)$ and
a model likelihood $p(d|\theta)$, the Bayesian posterior
\begin{equation}\label{eq:Bayes}
  p(\theta|d) = \frac{p(d|\theta)p(\theta)}{p(d)}
\end{equation}
summarises, as a probability distribution, our knowledge of the model parameters $\theta$ 
after observing data $d$. 
 When $p(d|\theta)$ is tractable (as in the case of
GWs) likelihood-based samplers such as Markov chain Monte Carlo
(MCMC)~\cite{metropolis1953equation,Hastings:1970} or nested
sampling~\cite{Skilling:2006} are typically used to draw samples from
the posterior. If it is possible to \emph{sample} $d\sim p(d|\theta)$
(i.e., simulate data) one can alternatively use amortized
simulation-based (or likelihood-free) inference
methods~\citep{cranmer2020frontier}. These approaches are based on
deep neural networks and can be several orders-of-magnitude faster at
inference time. For GW inference, they have also been shown to achieve
similar accuracy to MCMC~\cite{Dax:2021tsq}. In general, however, it
is not clear how well such networks generalize to out-of-distribution
data and they lack diagnostics to be confident in
results~\cite{cannon2022investigating}. These powerful approaches are
therefore rarely used in applications where accuracy is important and
likelihoods are tractable.

In this Letter, we achieve the best of both worlds by combining
likelihood-free and likelihood-based methods for GW parameter
estimation. We take samples from \textsc{Dingo}\footnote{Deep
  INference for Gravitational-wave
  Observations.}~\cite{Dax:2021tsq}---a fast and accurate
likelihood-free method using normalizing
flows~\cite{rezende2015variational,kingma2016improved,durkan2019neural,papamakarios2021normalizing}---and
treat these as a proposal for importance
sampling~\cite{tokdar2010importance}. The combined method
(``\textsc{Dingo-IS}'') 
generates samples from the exact posterior and now provides 
an estimate of the Bayesian evidence $p(d)$. Moreover, the importance
sampling efficiency arises as a powerful and objective performance
metric, which flags potential failure cases. Importance sampling is fully parallelizable.

After describing the method more fully in the following section, we
verify on two real events that \textsc{Dingo-IS} produces results
consistent with standard inference
codes~\cite{Veitch:2014wba,Ashton:2018jfp,Romero-Shaw:2020owr,Speagle_2020}. Our
main result is an analysis of 42 events from the Second and Third
Gravitational-Wave Transient Catalogs (GWTC-2 and
GWTC-3)~\cite{LIGOScientific:2020ibl,LIGOScientific:2021djp}, using
two waveform models, IMRPhenomXPHM~\cite{Pratten:2020ceb} and
SEOBNRv4PHM~\citep{Ossokine:2020kjp}. Due to the long waveform
simulation times, SEOBNRv4PHM inference would take several months per
event with stochastic samplers. However \textsc{Dingo-IS} with 64 CPU
cores takes just 10 hours for these waveforms. (Initial \textsc{Dingo}
samples are available typically in under a minute.)  Our results
indicate that \textsc{Dingo(-IS)} performs well for the majority of
events, and that failure cases are indeed flagged by low sample
efficiency. We also find that the log evidence is recovered with
statistical uncertainty reduced by a factor of 10 compared to standard
samplers.

Machine learning methods have seen numerous applications in GW astronomy, including to detection and parameter estimation~\cite{Cuoco:2020ogp}. For parameter estimation, these methods have included variational inference~\cite{Gabbard:2019rde,Green:2020hst}, likelihood ratio estimation~\cite{Delaunoy:2020zcu}, and posterior estimation with normalizing flows~\cite{Green:2020hst,Green:2020dnx,Dax:2021tsq,Chatterjee:2022ggk}. Aside from directly estimating parameters, normalizing flows have also been used to accelerate classical samplers, with significant efficiency improvements~\cite{Williams:2021qyt}.

Neural density estimation and importance sampling have previously been
combined~\cite{paige2016inference} under the guise of ``neural importance
sampling''~\cite{muller2019neural}, and similar approaches have been applied in several contexts~\cite{noe2019boltzmann,Albergo:2019eim,Kanwar:2020xzo,sun2022alpha}. Our contributions are
to (1) extend this to amortized simulation-based inference, (2) use it to improve results generated with
classical inference methods such as MCMC, and (3) to highlight how the
use of a forward Kullback-Leibler (KL) loss improves reliability. We
also apply it to the challenging real-world problem of GW
inference.\footnote{A similar approach using
  convolutional networks to parametrize Gaussian and von Mises
  proposals was used to estimate the sky position
  alone~\cite{Kolmus:2021buf} Using the normalizing flow proposal (as
  we do here) significantly improves the flexiblity of the conditional
  density estimator and enables inference of all parameters.} 
We demonstrate results that far outperform classical methods in terms of
sample efficiency and parallelizability, while maintaining accuracy and including simple diagnostics. We therefore expect this
work to accelerate the development and verification of probabilistic deep
learning approaches across science.

\sec{Method}\textsc{Dingo} trains a conditional density-estimation neural
network $q(\theta|d)$ to approximate $p(\theta|d)$ based on simulated
data sets $(\theta,d)$ with $\theta\sim p(\theta)$,
$d\sim p(d|\theta)$---an approach called neural posterior estimation
(NPE)~\cite{papamakarios2016fast}. Once trained, \textsc{Dingo} can
rapidly produce (approximate) posterior samples for any measured data
$d$. In practice, results may deviate from the true posterior due to
insufficient training, lack of network expressivity, or
out-of-distribution (OOD) data (i.e., data inconsistent with the
training distribution). Although it was shown in~\cite{Dax:2021tsq}
that these deviations are often negligible, verification of
results requires comparing against expensive standard samplers.

Here, we describe an efficient method to \emph{verify} and
\emph{correct} \textsc{Dingo} results using importance sampling
(IS)~\cite{tokdar2010importance}. Starting from a collection of $n$
samples $\theta_i\sim q(\theta|d)$ (the ``proposal'') we assign to
each one an importance weight
$w_i = p(d|\theta_i) p(\theta_i)/q(\theta_i|d)$. For a perfect
proposal, $w_i = \text{constant}$, but more generally the number of
\emph{effective samples} is related to the variance,
$n_\text{eff} = (\sum_i w_i)^2/\sum_i (w_i^2)$~\cite{kong1992note}. The \emph{sample
  efficiency} $\epsilon = n_\text{eff}/n \in (0,1]$ arises naturally
as a quality measure of the proposal.

Importance sampling requires evaluation of $p(d|\theta)p(\theta)$ rather
than the normalized posterior. The Bayesian evidence can then be
estimated from the normalization of the weights as
$p(d) = 1/n \sum_i w_i$. The standard deviation of the log evidence, $\sigma_{\log p(d)} = \sqrt{(1 - \epsilon)/(n\cdot \epsilon)}$ (see
Supplemental Material), scales with $1/\sqrt{n}$, enabling very
precise estimates. The evidence is furthermore unbiased if the support
of the posterior is fully covered by the proposal
distribution~\cite{mcbook}. The \emph{log} evidence does have a bias, but this scales as $1/n$, and in all cases considered here 
is completely negligible (see Supplemental Material). 
If $q(\theta|d)$ fails to cover the entire
posterior, the evidence itself would also be biased, toward lower values. 

NPE is particularly well-suited for IS because of two key properties. First, by construction the proposal has tractable density, such that we can not only sample from $q(\theta|d)$, but also evaluate it. 
Second, the NPE proposal is expected to always cover the entire posterior support. This is because, during training, NPE minimizes the \emph{forward} KL divergence $D_{\text{KL}}(p(\theta|d)||q(\theta|d))$. This diverges unless $\text{supp}(p(\theta|d))\subseteq \text{supp}(q(\theta|d))$, making the loss ``probability-mass covering''. 
Probability mass coverage is not guaranteed for finite sets of samples generated with stochastic samplers like MCMC (which can miss distributional modes), or machine learning methods with other training objectives like variational inference~\cite{jordan1999introduction,wainwright2008graphical,rezende2015variational}.

Neural importance sampling can in fact be used to improve posterior
samples from \emph{any} inference method provided the likelihood is
tractable. If the method provides only samples (without density) then
one must first train an (unconditional) density estimator $q(\theta)$
(e.g., a normalizing
flow~\cite{rezende2015variational,kingma2016improved,papamakarios2017masked})
to use as proposal. This is generally fast for an unconditional flow,
and using the forward KL loss guarantees that the proposal will cover
the samples. Success, however, relies on the quality of the initial
samples: if they are light-tailed, sample efficiency will be poor, and
if they are not mass-covering, the evidence will be
biased. Nevertheless, for initial samples that well represent the
posterior, this technique can provide quick verification and improvement.

In the context of GWs, we refer to neural importance sampling with
\textsc{Dingo} as \textsc{Dingo-IS}. Although this technique requires
likelihood evaluations at inference time, in practice it is much
faster than other likelihood-based methods because of its high sample
efficiency and parallelizability. Indeed, \textsc{Dingo} samples are
independent and identically distributed, trivially enabling
full parallelization of likelihood evaluations. This is a crucial
advantage compared to inherently sequential methods such as MCMC.

\sec{Results}For our experiments, we prepare \textsc{Dingo}
networks as described in~\cite{Dax:2021tsq}, with several
modifications. First, we extend the priors over component masses to
$m_1, m_2 \in [10,120]~\mathrm{M}_\odot$ and dimensionless spin
magnitudes to $a_1, a_2 \in [0,0.99]$. We also use the waveform models
IMRPhenomXPHM~\cite{Pratten:2020ceb} and
SEOBNRv4PHM~\cite{Ossokine:2020kjp}, which include higher radiative
multipoles and more realistic precession. Finally, in addition to
networks for the first observing run of LIGO and Virgo (O1), we also train networks based on O3 noise. For the O3
analyses, we found performance improved by training separate
\textsc{Dingo} models with distance priors $[0.1,3]$~Gpc,
$[0.1,6]$~Gpc and $[0.1,12]$~Gpc. We continue to use frequency-domain
strain data in the range $[20, 1024]$~Hz with $\Delta f=0.125$~Hz and
identical data conditioning as in~\cite{Dax:2021tsq}. 
The network architecture, hyperparameters, and training algorithm are also unchanged. We consider the two LIGO~\cite{LIGOScientific:2014pky} detectors for all analyses, and leave inclusion of Virgo~\cite{VIRGO:2014yos} data to a future publication of a complete catalog.

In our experiments, we found that \textsc{Dingo} often has difficulty resolving the phase parameter $\phi_\text{c}$. Although $\phi_\text{c}$ itself is of little physical interest, it is nevertheless needed to evaluate the likelihood for importance sampling. We therefore sample $\phi_\text{c}$ synthetically, by first evaluating the likelihood across a $\phi_\text{c}$ grid and caching the waveform modes for efficiency (see Supplemental Material). This approach is similar to standard phase marginalization~\cite{veitch2013analytic,Veitch:2014wba,Thrane:2018qnx}, but it is valid even with higher modes; it can therefore be adapted also to stochastic samplers.

\begin{table}[]
  \centering
  \begin{tabular}{cccc}
    \toprule
    & ~~Mean JSD~~ & ~~Max JSD~~ & $\log p(d)$ \\\hline
    \textsc{Dingo} & 2.2 & 7.2 ($\alpha$) & -\\
    \textsc{Dingo}-IS & 0.5 & 1.4 ($d_\text{L}$) & $-15831.87\pm 0.01$\\
    \textsc{Bilby} & 1.8 & 4.0 ($d_\text{L}$) & $-15831.78\pm 0.10$\\\hline
    \textsc{Dingo} & 9.0 & 53.4 ($M_\text{c}$) & -\\
    \textsc{Dingo}-IS & 0.7 & 2.2 ($\alpha$) & $-16412.88\pm 0.01$\\
    \textsc{Bilby} & 1.1 & 4.1 ($\alpha$) & $-16412.73\pm 0.09$\\\toprule
  \end{tabular}
  \caption{Performance for GW150914 (upper block) and
    GW151012 (lower) with waveform model IMRPhenomXPHM. The Jensen-Shannon divergence (JSD) quantifies
    the deviation from \textsc{LALInference-MCMC} for one-dimensional
    marginal posteriors (all values in $10^{-3}$~nat). The mean is
    taken across all parameters. Posteriors with a maximum JSD
    $\leq 2 \times 10^{-3}$~nat are considered
    indistinguishable~\cite{Romero-Shaw:2020owr}; here, maxima occur
    for right ascension $\alpha$, luminosity distance $d_\text{L}$, and chirp
    mass $M_\text{c}$. 
    We also report \textsc{Bilby-dynesty} results.
    }
  \label{tab:O1_performance}
\end{table}

\begin{figure}
  \includegraphics[width=0.48\textwidth]{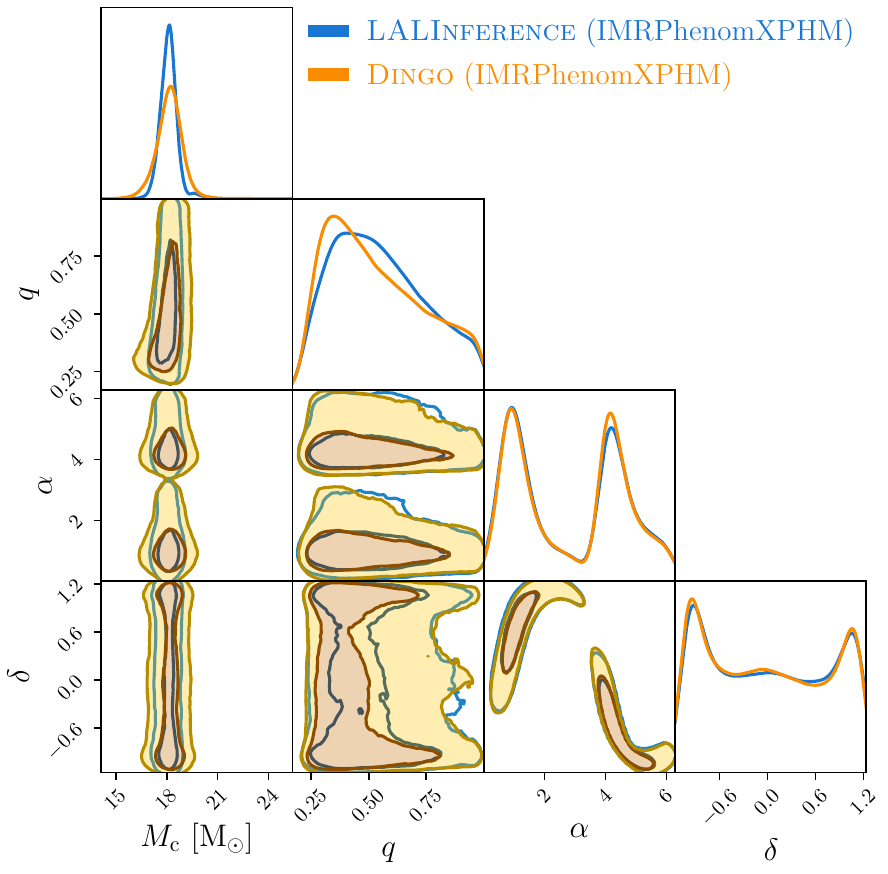}
  \hfill
  \includegraphics[width=0.48\textwidth]{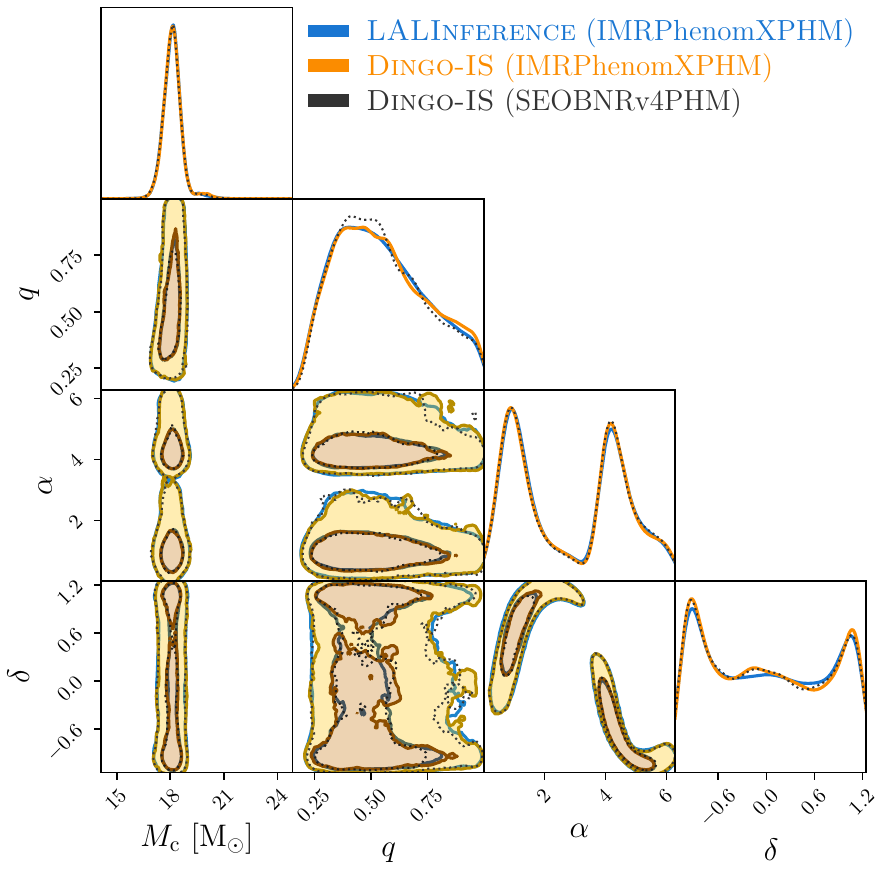}
  \caption{\label{fig:GW151012-XPHM}Chirp mass ($M_\text{c}$), mass ratio
    ($q$) and sky position ($\alpha,\delta$) parameters for GW151012,
    comparing inference with \textsc{Dingo} and
    \textsc{LALInference-MCMC}. Even when initial \textsc{Dingo}
    results deviate from \textsc{LALInference} posteriors (upper
    panel), IS leads to almost perfect agreement (lower). For
    comparison, the lower panel also shows results for SEOBNRv4PHM.}
\end{figure}

\newcommand{\GrayRow}{\cellcolor{lightgray}}

\begin{table*}[]
  \centering
  \begin{tabular}{lcr}\hline\hline
  Event & $\log p(d)$ &  \multicolumn{1}{c}{$\epsilon$}\\\hline\hline
    GW190408 & $-16178.332\pm 0.012$ & \cellcolor[HTML]{D1EBB4} $6.9$\%\\
    \_181802 & $-16178.172\pm 0.010$ & \cellcolor[HTML]{C0E4B3} $9.3$\%\\\hline
    GW190413 & $-15571.413\pm 0.006$ & \cellcolor[HTML]{8CCAA7} $22.5$\%\\
    \_052954 & $-15571.391\pm 0.005$ & \cellcolor[HTML]{88C3A3} $26.3$\%\\\hline
    \GrayRow GW190413 & $-16399.331\pm 0.009$ & \cellcolor[HTML]{B1DDB1} $12.4$\%\\
    \GrayRow \_134308 & $-16399.139\pm 0.014$ & \cellcolor[HTML]{E1F2BE} $4.7$\%\\\hline
    GW190421 & $-15983.248\pm 0.008$ & \cellcolor[HTML]{A2D6AD} $15.3$\%\\
    \_213856 & $-15983.131\pm 0.010$ & \cellcolor[HTML]{BFE3B2} $9.4$\%\\\hline
    \GrayRow GW190503 & $-16582.865\pm 0.022$ & \cellcolor[HTML]{FFFFDE} $2.0$\%\\
    \GrayRow \_185404 & $-16583.352\pm 0.027$ & \cellcolor[HTML]{FFF5CE} $1.4$\%\\\hline
    \GrayRow GW190513 & $-15946.462\pm 0.043$ & \cellcolor[HTML]{FED3AF} $0.6$\%\\
    \GrayRow \_205428 & $-15946.581\pm 0.017$ & \cellcolor[HTML]{EEF8C8} $3.4$\%\\\hline
    \GrayRow GW190514 & $-16556.466\pm 0.009$ & \cellcolor[HTML]{B4DFB1} $11.6$\%\\
    \GrayRow \_065416 & $-16556.314\pm 0.017$ & \cellcolor[HTML]{EDF8C7} $3.5$\%\\\hline
    GW190517 & $-16271.048\pm 0.027$ & \cellcolor[HTML]{FFF4CD} $1.3$\%\\
    \_055101 & $-16272.428\pm 0.034$ & \cellcolor[HTML]{FEE6BD} $0.9$\%\\\hline
    GW190519 & $-15991.171\pm 0.008$ & \cellcolor[HTML]{A4D7AD} $15.2$\%\\
    \_153544 & $-15991.287\pm 0.068$ & \cellcolor[HTML]{F0A298} $0.2$\%\\\hline
    GW190521 & $-16008.876\pm 0.008$ & \cellcolor[HTML]{ABDAAF} $13.4$\%\\
    \_074359 & $-16008.037\pm 0.015$ & \cellcolor[HTML]{E6F4C1} $4.2$\%\\\hline
    GW190527 & $-16119.012\pm 0.008$ & \cellcolor[HTML]{AADAAF} $13.8$\%\\
    \_092055 & $-16118.781\pm 0.013$ & \cellcolor[HTML]{D6EEB7} $6.1$\%\\\hline
    GW190602 & $-16036.993\pm 0.006$ & \cellcolor[HTML]{8AC6A5} $25.0$\%\\
    \_175927 & $-16037.529\pm 0.006$ & \cellcolor[HTML]{8BC8A6} $23.5$\%\\\hline
    \GrayRow GW190701 & $-16521.381\pm 0.040$ & \cellcolor[HTML]{FED8B1} $0.6$\%\\
    \GrayRow \_203306 & $-16521.609\pm 0.010$ & \cellcolor[HTML]{BBE2B2} $10.1$\%\\\hline
    GW190719 & $-15850.492\pm 0.008$ & \cellcolor[HTML]{ABDAAF} $13.4$\%\\
    \_215514 & $-15850.339\pm 0.011$ & \cellcolor[HTML]{C8E7B3} $8.0$\%\\\hline
  \end{tabular}\hspace{0.05cm}
  \begin{tabular}{lcr}\hline\hline
  Event & $\log p(d)$ &  \multicolumn{1}{c}{$\epsilon$}\\\hline\hline
    \GrayRow GW190727 & $-15992.017\pm 0.009$ & \cellcolor[HTML]{BBE2B2} $10.3$\%\\
    \GrayRow \_060333 & $-15992.428\pm 0.005$ & \cellcolor[HTML]{85BEA0} $30.8$\%\\\hline
    GW190731 & $-16376.777\pm 0.005$ & \cellcolor[HTML]{84BB9F} $32.6$\%\\
    \_140936 & $-16376.763\pm 0.005$ & \cellcolor[HTML]{85BDA0} $31.0$\%\\\hline
    GW190803 & $-16132.409\pm 0.006$ & \cellcolor[HTML]{8DCCA8} $21.4$\%\\
    \_022701 & $-16132.408\pm 0.005$ & \cellcolor[HTML]{87C2A2} $27.8$\%\\\hline
    GW190805 & $-16073.261\pm 0.006$ & \cellcolor[HTML]{92CEA9} $20.0$\%\\
    \_211137 & $-16073.656\pm 0.007$ & \cellcolor[HTML]{9ED4AC} $16.6$\%\\\hline
    GW190828 & $-16137.220\pm 0.009$ & \cellcolor[HTML]{B1DDB1} $12.2$\%\\
    \_063405 & $-16136.799\pm 0.010$ & \cellcolor[HTML]{C2E5B3} $9.1$\%\\\hline
    GW190909 & $-16061.634\pm 0.011$ & \cellcolor[HTML]{CDEAB4} $7.4$\%\\
    \_114149 & $-16061.275\pm 0.016$ & \cellcolor[HTML]{EBF7C4} $3.8$\%\\\hline
    GW190915 & $-16083.960\pm 0.015$ & \cellcolor[HTML]{8FCDA8} $20.8$\%\\
    \_235702 & $-16083.937\pm 0.027$ & \cellcolor[HTML]{E0F2BD} $4.8$\%\\\hline
    GW190926 & $-16015.813\pm 0.019$ & \cellcolor[HTML]{F4FAD0} $2.8$\%\\
    \_050336 & $-16015.861\pm 0.009$ & \cellcolor[HTML]{B2DEB1} $12.1$\%\\\hline
    GW190929 & $-16146.666\pm 0.018$ & \cellcolor[HTML]{F0F9CB} $3.2$\%\\
    \_012149 & $-16146.591\pm 0.021$ & \cellcolor[HTML]{F9FDD7} $2.4$\%\\\hline
    \GrayRow GW191109 & $-17925.064\pm 0.025$ & \cellcolor[HTML]{FFFAD6} $1.7$\%\\
    \GrayRow \_010717 & $-17922.762\pm 0.041$ & \cellcolor[HTML]{FED7B0} $0.6$\%\\\hline
    \GrayRow GW191127 & $-16759.328\pm 0.019$ & \cellcolor[HTML]{F6FBD2} $2.7$\%\\
    \GrayRow \_050227 & $-16758.102\pm 0.029$ & \cellcolor[HTML]{FFF1C8} $1.2$\%\\\hline
    \ddag GW191204 & $-15984.455\pm 0.015$ & \cellcolor[HTML]{E6F4C1} $4.2$\%\\
    \_110529 & $-15983.618\pm 0.063$ & \cellcolor[HTML]{F3A89A} $0.3$\%\\\hline
    GW191215 & $-16001.286\pm 0.013$ & \cellcolor[HTML]{D8EEB8} $5.8$\%\\
    \_223052 & $-16000.846\pm 0.052$ & \cellcolor[HTML]{FABCA4} $0.4$\%\\\hline
    GW191222 & $-15871.521\pm 0.007$ & \cellcolor[HTML]{9ED4AC} $16.5$\%\\
    \_033537 & $-15871.450\pm 0.005$ & \cellcolor[HTML]{89C4A4} $25.8$\%\\\hline
  \end{tabular}\hspace{0.05cm}
  \begin{tabular}{lcr}\hline\hline
  Event & $\log p(d)$ &  \multicolumn{1}{c}{$\epsilon$}\\\hline\hline
    GW191230 & $-15913.798\pm 0.009$ & \cellcolor[HTML]{B1DDB1} $12.2$\%\\
    \_180458 & $-15913.918\pm 0.010$ & \cellcolor[HTML]{C3E5B3} $8.8$\%\\\hline
    GW200128 & $-16305.128\pm 0.013$ & \cellcolor[HTML]{D6EEB7} $6.1$\%\\
    \_022011 & $-16304.510\pm 0.007$ & \cellcolor[HTML]{98D1AA} $18.3$\%\\\hline
    \GrayRow \ddag GW200129 & $-16226.851\pm 0.109$ & \cellcolor[HTML]{D28092} $0.1$\%\\
    \GrayRow \_065458 & $-16231.203\pm 0.051$ & \cellcolor[HTML]{FBBFA5} $0.4$\%\\\hline
    GW200208 & $-16136.381\pm 0.007$ & \cellcolor[HTML]{9ED4AC} $16.6$\%\\
    \_130117 & $-16136.531\pm 0.009$ & \cellcolor[HTML]{B6E0B1} $11.2$\%\\\hline
    GW200208 & $-16775.200\pm 0.011$ & \cellcolor[HTML]{CDEAB4} $7.4$\%\\
    \_222617 & $-16774.582\pm 0.021$ & \cellcolor[HTML]{FCFEDA} $2.2$\%\\\hline
    GW200209 & $-16383.847\pm 0.009$ & \cellcolor[HTML]{B0DDB0} $12.5$\%\\
    \_085452 & $-16384.157\pm 0.025$ & \cellcolor[HTML]{FFFAD6} $1.6$\%\\\hline
    GW200216 & $-16215.703\pm 0.017$ & \cellcolor[HTML]{EEF8C8} $3.4$\%\\
    \_220804 & $-16215.540\pm 0.018$ & \cellcolor[HTML]{F1F9CC} $3.1$\%\\\hline
    GW200219 & $-16133.457\pm 0.011$ & \cellcolor[HTML]{BEE3B2} $9.6$\%\\
    \_094415 & $-16133.157\pm 0.017$ & \cellcolor[HTML]{E8F5C2} $4.0$\%\\\hline
    GW200220 & $-16303.782\pm 0.007$ & \cellcolor[HTML]{9BD2AB} $17.3$\%\\
    \_061928 & $-16303.087\pm 0.026$ & \cellcolor[HTML]{FFF7D2} $1.5$\%\\\hline
    GW200220 & $-16136.600\pm 0.008$ & \cellcolor[HTML]{ADDBB0} $13.2$\%\\
    \_124850 & $-16136.519\pm 0.037$ & \cellcolor[HTML]{FEDFB7} $0.7$\%\\\hline
    GW200224 & $-16138.613\pm 0.006$ & \cellcolor[HTML]{8CCAA7} $22.5$\%\\
    \_222234 & $-16139.101\pm 0.006$ & \cellcolor[HTML]{8DCCA8} $21.4$\%\\\hline
    \ddag GW200308 & $-16173.938\pm 0.013$ & \cellcolor[HTML]{D6EEB7} $6.0$\%\\
    \_173609 & $-16173.692\pm 0.025$ & \cellcolor[HTML]{FFFAD7} $1.7$\%\\\hline
    GW200311 & $-16117.505\pm 0.011$ & \cellcolor[HTML]{CDEAB4} $7.4$\%\\
    \_115853 & $-16117.583\pm 0.009$ & \cellcolor[HTML]{B2DEB1} $11.9$\%\\\hline
    \ddag GW200322 & $-16313.568\pm 0.307$ & \cellcolor[HTML]{D28092} $0.0$\%\\
    \_091133 & $-16313.110\pm 0.105$ & \cellcolor[HTML]{D28092} $0.1$\%\\\hline
  \end{tabular}
  \caption{42 BBH events from GWTC-3 analyzed with
    \textsc{Dingo-IS}. We report the log evidence $\log p(d)$ and the
    sample efficiency $\epsilon$ for the two waveform models
    IMRPhenomXPHM (upper rows) and SEOBNRv4PHM (lower
    rows). Highlighting colors indicate the sample efficiency (green:
    high; yellow: medium; orange/red: low); \textsc{Dingo-IS} results
    can be trusted for medium and high $\epsilon$ (see Supplemental
    Material). Events in gray suffer from data quality
    issues~\cite{LIGOScientific:2020ibl,LIGOScientific:2021djp}.
    \ddag See remarks on these events in text.}
  \label{tab:O3_events}
\end{table*}

For \textsc{Dingo-IS}, with $10^5$ proposal samples per event, the
total time for inference using one NVIDIA A100 GPU and 64 CPU cores is
typically less than 1~hour for IMRPhenomXPHM 
and $\approx 10$~hours for SEOBNRv4PHM. In both cases, the computation time is  
dominated by waveform
simulations, which could be further reduced
using more CPUs. The rest of the time is taken up to generate the initial \textsc{Dingo}
proposal samples.\footnote{It takes longer to generate the proposal than to produce low-latency \textsc{Dingo} samples ($\approx 20~\text{s}$) because of the group-equivariant NPE (GNPE) algorithm~\cite{Dax:2021tsq,Dax:2021myb} (which breaks access to the density) and the synthetic phase recovery. See Supplemental Material for details.}

We first validate \textsc{Dingo-IS} against standard inference codes
for two real events, GW150914 and GW151012, using IMRPhenomXPHM. 
(For SEOBNRv4PHM it is not feasible to run classical samplers, and one would instead need to use faster methods such as \textsc{RIFT}~\cite{Pankow:2015cra,Lange:2018pyp}.) 
We generate reference posteriors using
\textsc{LALInference-MCMC}~\cite{Veitch:2014wba}, and compare
one-dimensional marginalized posteriors for each parameter using the
Jensen-Shannon divergence (Tab.~\ref{tab:O1_performance}). For both
events, the initial small deviations of \textsc{Dingo}
samples from the reference are made negligible\footnote{
    Initial deviations are larger than those reported in~\cite{Dax:2021tsq} since we use a more complicated waveform model and a larger prior, while keeping the size of the neural network and training time the same. Any remaining deviations after importance sampling can in principle also be due to sampling inaccuracies of \textsc{LALInference-MCMC}. Note that a direct comparison to published LIGO-Virgo-KAGRA results is impeded by different data settings.
} using \textsc{Dingo-IS}
(see Fig.~\ref{fig:GW151012-XPHM} for a qualitative demonstration). 
We
find sample efficiencies of $\epsilon = 28.8\%$ and $\epsilon=12.5\%$
for GW150914 and GW151012, respectively.

For the evidence, we compare against
\textsc{Bilby-dynesty}~\cite{Ashton:2018jfp,Romero-Shaw:2020owr,Speagle_2020},
since nested sampling generally provides a more accurate estimate than
MCMC. In Tab.~\ref{tab:O1_performance} we see that \textsc{Dingo-IS}
is more precise by a factor of $\approx 10$, but the \textsc{Bilby}
evidence is larger for both events by roughly one standard
deviation. This deviation could be statistical, but it could also
indicate a bias in one of the methods. (Recall that IS requires the
proposal to be mass-covering for an unbiased evidence.)  To further
investigate for GW151012, we perform neural importance sampling
starting from $10^6$ \textsc{Bilby} samples (see Supplemental
Material). This achieves a slightly lower $\epsilon = 8.3\%$ than
\textsc{Dingo-IS}, but $\log p(d) = -16412.89\pm 0.01$ in close
agreement. While this does not fully rule out a bias in
\textsc{Dingo-IS} samples (since the test is not fully independent) we
take this as an indication that \textsc{Dingo-IS} indeed infers an
unbiased evidence.
More generally, it showcases how our method can be extended to improve the output of stochastic samplers. 

We now perform a large study analyzing all 42 events in
GWTC-2~\cite{LIGOScientific:2020ibl} and
GWTC-3~\cite{LIGOScientific:2021djp} that are consistent with our mass prior.\footnote{\label{footnote:bns}Lower mass events produce longer signals, so extending \textsc{Dingo} to these may require improved methods for data compression~\cite{Vinciguerra:2017ngf,Cannon:2011vi}. This will be particularly relevant for binary neutron stars.} 
We stress
that a study of this scope would be infeasible with standard codes,
since SEOBNRv4PHM inference for a single event would take several
months. Across all events we achieve a median sampling efficiency of
$\epsilon = 10.9\%$ for IMRPhenomXPHM and $\epsilon = 4.4\%$ for
SEOBNRv4PHM (Tab.~\ref{tab:O3_events}). For most events, the initial
\textsc{Dingo} results are already accurate and only deviate slightly
from \textsc{Dingo-IS}; furthermore, \textsc{Dingo-IS} shows excellent agreement between the two waveform models (see the Supplemental Material for more detailed comparisons).  
Note that these results are based on highly complex precessing higher-mode waveform models, and do not include any mitigation of noise transients (see below). With the simpler IMRPhenomPv2~\cite{Hannam:2013oca,Khan:2015jqa,Bohe:2016} model and a smaller mass prior (in a study on drifting detector noise distributions~\cite{wildberger2022probabilistic}) \textsc{Dingo-IS} achieves an even larger median sample efficiency of $\epsilon = 36.8\%$ on 37 events.

Importance sampling guarantees robust results by marking failure cases
with a low sample efficiency. By this metric, \textsc{Dingo} struggles
slightly with chirp masses near the lower prior boundary
(GW191204\_110529 and GW200322\_091133). For such systems, efficiency
may be improved by increasing the prior range used for
training. Events with known data quality issues also often have low sample efficiency (see Tab.~\ref{tab:O3_events}): several low-$\epsilon$ events are contaminated by glitch artifacts (which would be mitigated in a more complete analysis~\cite{LIGOScientific:2020ibl,LIGOScientific:2021djp}); GW200129\_065458, in addition to having a glitch~\cite{Payne:2022spz}, may not be well modeled by either of our waveform models due to having strong precession~\cite{Hannam:2021pit}; and GW200322\_091133 may be simply a Gaussian noise fluctuation~\cite{Morras:2022ysx}. In these cases, \textsc{Dingo-IS} marks events for additional investigation.

Data quality issues such as non-Gaussian noise or observed signals that do not match models correspond to OOD data, i.e., data not consistent with the training distribution. Since OOD data are not seen during training, \textsc{Dingo} cannot be expected to return their true posterior, which results in a low sample efficiency. As an additional test, running \textsc{Dingo-IS} on signal-free data with a blip glitch~\cite{coughlin_scott_2021_5649212} in the LIGO Hanford detector (GPS time 1238613687.5) results in $\epsilon\approx 0.001\%$. Likewise, we find that \textsc{Dingo-IS} successfully flags adversarial examples~\cite{szegedy2013intriguing,goodfellow2014explaining} that are intentionally corrupted to mislead the inference network ($\epsilon\approx 0.01\%$; see Supplemental Material)---addressing a common failure mode of neural networks. Our general view, therefore, is that although there can be various reasons for low-$\epsilon$ results, it often serves as a useful heuristic to identify OOD events.

\sec{Conclusions}We have described the use of importance sampling to
improve the results of NPE in amortized inference problems, and we
applied it to the case of GWs. Neural importance sampling provides
rapid verification of results and corrects any inaccuracies in deep
learning output; it provides an evidence estimate with precision far exceeding that
of classical samplers; and it marks potentially OOD data for
further investigation. With high sample efficiency and rapid initial
results, \textsc{Dingo-IS} becomes a comprehensive inference tool for
accurately analyzing the large numbers of BBH events expected soon.

High sample efficiencies are predicated on a high quality proposal,
which \textsc{Dingo} thankfully provides. A key element is the
probability-mass covering property, which is guaranteed by the forward
KL training loss. This tends to produce broad tails, which are
downweighted in importance sampling. \emph{Overly} broad proposals
would nevertheless result in low sample efficiency, so highly
expressive density estimators such as normalizing flows are essential,
along with \textsc{Dingo} innovations such as GNPE and GW training
data augmentation. \textsc{Dingo} posteriors are rarely light tailed,
but this does occasionally lead to underestimated evidence for small
$n$. 

With the inclusion of importance sampling, the \textsc{Dingo} pipeline
can now be used in several different ways. When low latency is
desired, complete posteriors are still available without importance
sampling in a matter of seconds. Results include sky position and mass
parameters and could therefore play an important role in directing
electromagnetic followup observations once we extend \textsc{Dingo} to mergers involving neutron stars (see footnote \ref{footnote:bns}). By comparing against
\textsc{Dingo-IS}, we have shown that in the majority of cases,
initial results are already very reliable, with only minor deviations
in marginal distributions. Indeed, validation of \textsc{Dingo}
results was a major motivation in exploring importance sampling.

When high accuracy is desired, \textsc{Dingo-IS} reweights results to
the true posterior and includes an estimate of the evidence. Results
are verified and include probability mass-covering guarantees that
ensure secondary modes are not missed.  Sample efficiencies are often
two orders-of-magnitude higher than MCMC or nested sampling, and
importance sampling is fully parallelizable. As a consequence, results
are typically available within an hour for IMRPhenomXPHM, or 10 hours
for SEOBNRv4PHM. This represents a significant advantage when
considering the event rates likely to be reached with advanced
detectors (three per week or higher in the upcoming LIGO-Virgo-KAGRA observing run O4).

\textsc{Dingo-IS} opens several new possibilities for GW analysis: 
(1) rapid inference means that the most accurate waveform models, which include all physical effects, could be used for all events;
(2) high-precision evidences enable detailed
model comparison; and (3) low sample efficiencies can identify data
that do not fit the noise or waveform model. We believe that these
results have highlighted clear benefits of combining likelihood-free
and likelihood-based methods in Bayesian inference. Going forward, as \textsc{Dingo-IS}
validates and builds trust in \textsc{Dingo}, it will help to set the
stage for noise-model free inference, which is truly likelihood-free.

The code for \textsc{Dingo} and \textsc{Dingo-IS} is available at \href{https://github.com/dingo-gw/dingo}{https://github.com/dingo-gw/dingo}.

\begin{acknowledgments}
\sec{Acknowledgments}We thank V. Raymond for encouraging us to pursue importance sampling in the early stages of the project, and C. Garc\'ia Quir\'os, N. Gupte, S. Ossokine, A. Ramos-Buades and R. Smith for useful discussions. 
This material is based upon work supported by
NSF's LIGO Laboratory which is a major facility fully funded by the
National Science Foundation. This research has made use of data or
software obtained from the Gravitational Wave Open Science Center
(gw-openscience.org), a service of LIGO Laboratory, the LIGO
Scientific Collaboration, the Virgo Collaboration, and KAGRA. LIGO
Laboratory and Advanced LIGO are funded by the United States National
Science Foundation (NSF) as well as the Science and Technology
Facilities Council (STFC) of the United Kingdom, the
Max-Planck-Society (MPS), and the State of Niedersachsen/Germany for
support of the construction of Advanced LIGO and construction and
operation of the GEO600 detector. Additional support for Advanced LIGO
was provided by the Australian Research Council. Virgo is funded,
through the European Gravitational Observatory (EGO), by the French
Centre National de Recherche Scientifique (CNRS), the Italian Istituto
Nazionale di Fisica Nucleare (INFN) and the Dutch Nikhef, with
contributions by institutions from Belgium, Germany, Greece, Hungary,
Ireland, Japan, Monaco, Poland, Portugal, Spain. The construction and
operation of KAGRA are funded by Ministry of Education, Culture,
Sports, Science and Technology (MEXT), and Japan Society for the
Promotion of Science (JSPS), National Research Foundation (NRF) and
Ministry of Science and ICT (MSIT) in Korea, Academia Sinica (AS) and
the Ministry of Science and Technology (MoST) in Taiwan. 
M.D. thanks the Hector Fellow Academy for support. J.H.M. and B.S. are members of the MLCoE, EXC number 2064/1 – Project number 390727645 and the Tübingen AI Center funded by the German Ministry for Science and Education (FKZ 01IS18039A).
For the implementation of \textsc{Dingo} we use \verb|PyTorch|~\cite{NEURIPS2019_9015}, \verb|nflows|~\cite{nflows}, \verb|LALSimulation|~\cite{lalsuite} and the adam optimizer~\cite{Kingma:2014vow}. The plots are generated with \verb|matplotlib|~\cite{Hunter:2007} and \verb|ChainConsumer|~\cite{Hinton2016}.
\end{acknowledgments}

\bibliography{mybib.bib}

\begin{thebibliography}{68}%
\makeatletter
\providecommand \@ifxundefined [1]{%
 \@ifx{#1\undefined}
}%
\providecommand \@ifnum [1]{%
 \ifnum #1\expandafter \@firstoftwo
 \else \expandafter \@secondoftwo
 \fi
}%
\providecommand \@ifx [1]{%
 \ifx #1\expandafter \@firstoftwo
 \else \expandafter \@secondoftwo
 \fi
}%
\providecommand \natexlab [1]{#1}%
\providecommand \enquote  [1]{``#1''}%
\providecommand \bibnamefont  [1]{#1}%
\providecommand \bibfnamefont [1]{#1}%
\providecommand \citenamefont [1]{#1}%
\providecommand \href@noop [0]{\@secondoftwo}%
\providecommand \href [0]{\begingroup \@sanitize@url \@href}%
\providecommand \@href[1]{\@@startlink{#1}\@@href}%
\providecommand \@@href[1]{\endgroup#1\@@endlink}%
\providecommand \@sanitize@url [0]{\catcode `\\12\catcode `\$12\catcode
  `\&12\catcode `\#12\catcode `\^12\catcode `\_12\catcode `\%12\relax}%
\providecommand \@@startlink[1]{}%
\providecommand \@@endlink[0]{}%
\providecommand \url  [0]{\begingroup\@sanitize@url \@url }%
\providecommand \@url [1]{\endgroup\@href {#1}{\urlprefix }}%
\providecommand \urlprefix  [0]{URL }%
\providecommand \Eprint [0]{\href }%
\providecommand \doibase [0]{https://doi.org/}%
\providecommand \selectlanguage [0]{\@gobble}%
\providecommand \bibinfo  [0]{\@secondoftwo}%
\providecommand \bibfield  [0]{\@secondoftwo}%
\providecommand \translation [1]{[#1]}%
\providecommand \BibitemOpen [0]{}%
\providecommand \bibitemStop [0]{}%
\providecommand \bibitemNoStop [0]{.\EOS\space}%
\providecommand \EOS [0]{\spacefactor3000\relax}%
\providecommand \BibitemShut  [1]{\csname bibitem#1\endcsname}%
\let\auto@bib@innerbib\@empty
\bibitem [{\citenamefont {Abbott}\ \emph
  {et~al.}(2021{\natexlab{a}})\citenamefont {Abbott} \emph
  {et~al.}}]{LIGOScientific:2021djp}%
  \BibitemOpen
  \bibfield  {author} {\bibinfo {author} {\bibfnamefont {R.}~\bibnamefont
  {Abbott}} \emph {et~al.} (\bibinfo {collaboration} {LIGO Scientific, VIRGO,
  KAGRA}),\ }\bibfield  {title} {\bibinfo {title} {{GWTC-3: Compact Binary
  Coalescences Observed by LIGO and Virgo During the Second Part of the Third
  Observing Run}},\ }\href@noop {} {\  (\bibinfo {year}
  {2021}{\natexlab{a}})},\ \Eprint {https://arxiv.org/abs/2111.03606}
  {arXiv:2111.03606 [gr-qc]} \BibitemShut {NoStop}%
\bibitem [{\citenamefont {Abbott}\ \emph
  {et~al.}(2021{\natexlab{b}})\citenamefont {Abbott} \emph
  {et~al.}}]{LIGOScientific:2021sio}%
  \BibitemOpen
  \bibfield  {author} {\bibinfo {author} {\bibfnamefont {R.}~\bibnamefont
  {Abbott}} \emph {et~al.} (\bibinfo {collaboration} {LIGO Scientific, VIRGO,
  KAGRA}),\ }\bibfield  {title} {\bibinfo {title} {{Tests of General Relativity
  with GWTC-3}},\ }\href@noop {} {\  (\bibinfo {year} {2021}{\natexlab{b}})},\
  \Eprint {https://arxiv.org/abs/2112.06861} {arXiv:2112.06861 [gr-qc]}
  \BibitemShut {NoStop}%
\bibitem [{\citenamefont {Abbott}\ \emph {et~al.}(2018)\citenamefont {Abbott}
  \emph {et~al.}}]{Abbott:2018exr}%
  \BibitemOpen
  \bibfield  {author} {\bibinfo {author} {\bibfnamefont {B.~P.}\ \bibnamefont
  {Abbott}} \emph {et~al.} (\bibinfo {collaboration} {LIGO Scientific,
  Virgo}),\ }\bibfield  {title} {\bibinfo {title} {{GW170817: Measurements of
  neutron star radii and equation of state}},\ }\href
  {https://doi.org/10.1103/PhysRevLett.121.161101} {\bibfield  {journal}
  {\bibinfo  {journal} {Phys. Rev. Lett.}\ }\textbf {\bibinfo {volume} {121}},\
  \bibinfo {pages} {161101} (\bibinfo {year} {2018})},\ \Eprint
  {https://arxiv.org/abs/1805.11581} {arXiv:1805.11581 [gr-qc]} \BibitemShut
  {NoStop}%
\bibitem [{\citenamefont {Abbott}\ \emph
  {et~al.}(2021{\natexlab{c}})\citenamefont {Abbott} \emph
  {et~al.}}]{LIGOScientific:2021psn}%
  \BibitemOpen
  \bibfield  {author} {\bibinfo {author} {\bibfnamefont {R.}~\bibnamefont
  {Abbott}} \emph {et~al.} (\bibinfo {collaboration} {LIGO Scientific, VIRGO,
  KAGRA}),\ }\bibfield  {title} {\bibinfo {title} {{The population of merging
  compact binaries inferred using gravitational waves through GWTC-3}},\
  }\href@noop {} {\  (\bibinfo {year} {2021}{\natexlab{c}})},\ \Eprint
  {https://arxiv.org/abs/2111.03634} {arXiv:2111.03634 [astro-ph.HE]}
  \BibitemShut {NoStop}%
\bibitem [{\citenamefont {Abbott}\ \emph
  {et~al.}(2021{\natexlab{d}})\citenamefont {Abbott} \emph
  {et~al.}}]{LIGOScientific:2021aug}%
  \BibitemOpen
  \bibfield  {author} {\bibinfo {author} {\bibfnamefont {R.}~\bibnamefont
  {Abbott}} \emph {et~al.} (\bibinfo {collaboration} {LIGO Scientific, VIRGO,
  KAGRA}),\ }\bibfield  {title} {\bibinfo {title} {{Constraints on the cosmic
  expansion history from GWTC-3}},\ }\href@noop {} {\  (\bibinfo {year}
  {2021}{\natexlab{d}})},\ \Eprint {https://arxiv.org/abs/2111.03604}
  {arXiv:2111.03604 [astro-ph.CO]} \BibitemShut {NoStop}%
\bibitem [{\citenamefont {Metropolis}\ \emph {et~al.}(1953)\citenamefont
  {Metropolis}, \citenamefont {Rosenbluth}, \citenamefont {Rosenbluth},
  \citenamefont {Teller},\ and\ \citenamefont
  {Teller}}]{metropolis1953equation}%
  \BibitemOpen
  \bibfield  {author} {\bibinfo {author} {\bibfnamefont {N.}~\bibnamefont
  {Metropolis}}, \bibinfo {author} {\bibfnamefont {A.~W.}\ \bibnamefont
  {Rosenbluth}}, \bibinfo {author} {\bibfnamefont {M.~N.}\ \bibnamefont
  {Rosenbluth}}, \bibinfo {author} {\bibfnamefont {A.~H.}\ \bibnamefont
  {Teller}},\ and\ \bibinfo {author} {\bibfnamefont {E.}~\bibnamefont
  {Teller}},\ }\bibfield  {title} {\bibinfo {title} {Equation of state
  calculations by fast computing machines},\ }\href@noop {} {\bibfield
  {journal} {\bibinfo  {journal} {The journal of chemical physics}\ }\textbf
  {\bibinfo {volume} {21}},\ \bibinfo {pages} {1087} (\bibinfo {year}
  {1953})}\BibitemShut {NoStop}%
\bibitem [{\citenamefont {Hastings}(1970)}]{Hastings:1970}%
  \BibitemOpen
  \bibfield  {author} {\bibinfo {author} {\bibfnamefont {W.~K.}\ \bibnamefont
  {Hastings}},\ }\bibfield  {title} {\bibinfo {title} {{Monte Carlo sampling
  methods using Markov chains and their applications}},\ }\href
  {https://doi.org/10.1093/biomet/57.1.97} {\bibfield  {journal} {\bibinfo
  {journal} {Biometrika}\ }\textbf {\bibinfo {volume} {57}},\ \bibinfo {pages}
  {97} (\bibinfo {year} {1970})},\ \Eprint
  {https://arxiv.org/abs/https://academic.oup.com/biomet/article-pdf/57/1/97/23940249/57-1-97.pdf}
  {https://academic.oup.com/biomet/article-pdf/57/1/97/23940249/57-1-97.pdf}
  \BibitemShut {NoStop}%
\bibitem [{\citenamefont {Skilling}(2006)}]{Skilling:2006}%
  \BibitemOpen
  \bibfield  {author} {\bibinfo {author} {\bibfnamefont {J.}~\bibnamefont
  {Skilling}},\ }\bibfield  {title} {\bibinfo {title} {{Nested sampling for
  general Bayesian computation}},\ }\href {https://doi.org/10.1214/06-BA127}
  {\bibfield  {journal} {\bibinfo  {journal} {Bayesian Analysis}\ }\textbf
  {\bibinfo {volume} {1}},\ \bibinfo {pages} {833 } (\bibinfo {year}
  {2006})}\BibitemShut {NoStop}%
\bibitem [{\citenamefont {Cranmer}\ \emph {et~al.}(2020)\citenamefont
  {Cranmer}, \citenamefont {Brehmer},\ and\ \citenamefont
  {Louppe}}]{cranmer2020frontier}%
  \BibitemOpen
  \bibfield  {author} {\bibinfo {author} {\bibfnamefont {K.}~\bibnamefont
  {Cranmer}}, \bibinfo {author} {\bibfnamefont {J.}~\bibnamefont {Brehmer}},\
  and\ \bibinfo {author} {\bibfnamefont {G.}~\bibnamefont {Louppe}},\
  }\bibfield  {title} {\bibinfo {title} {The frontier of simulation-based
  inference},\ }\href@noop {} {\bibfield  {journal} {\bibinfo  {journal}
  {Proceedings of the National Academy of Sciences}\ }\textbf {\bibinfo
  {volume} {117}},\ \bibinfo {pages} {30055} (\bibinfo {year}
  {2020})}\BibitemShut {NoStop}%
\bibitem [{\citenamefont {Dax}\ \emph {et~al.}(2021)\citenamefont {Dax},
  \citenamefont {Green}, \citenamefont {Gair}, \citenamefont {Macke},
  \citenamefont {Buonanno},\ and\ \citenamefont {Sch\"olkopf}}]{Dax:2021tsq}%
  \BibitemOpen
  \bibfield  {author} {\bibinfo {author} {\bibfnamefont {M.}~\bibnamefont
  {Dax}}, \bibinfo {author} {\bibfnamefont {S.~R.}\ \bibnamefont {Green}},
  \bibinfo {author} {\bibfnamefont {J.}~\bibnamefont {Gair}}, \bibinfo {author}
  {\bibfnamefont {J.~H.}\ \bibnamefont {Macke}}, \bibinfo {author}
  {\bibfnamefont {A.}~\bibnamefont {Buonanno}},\ and\ \bibinfo {author}
  {\bibfnamefont {B.}~\bibnamefont {Sch\"olkopf}},\ }\bibfield  {title}
  {\bibinfo {title} {{Real-Time Gravitational Wave Science with Neural
  Posterior Estimation}},\ }\href
  {https://doi.org/10.1103/PhysRevLett.127.241103} {\bibfield  {journal}
  {\bibinfo  {journal} {Phys. Rev. Lett.}\ }\textbf {\bibinfo {volume} {127}},\
  \bibinfo {pages} {241103} (\bibinfo {year} {2021})},\ \Eprint
  {https://arxiv.org/abs/2106.12594} {arXiv:2106.12594 [gr-qc]} \BibitemShut
  {NoStop}%
\bibitem [{\citenamefont {Cannon}\ \emph {et~al.}(2022)\citenamefont {Cannon},
  \citenamefont {Ward},\ and\ \citenamefont
  {Schmon}}]{cannon2022investigating}%
  \BibitemOpen
  \bibfield  {author} {\bibinfo {author} {\bibfnamefont {P.}~\bibnamefont
  {Cannon}}, \bibinfo {author} {\bibfnamefont {D.}~\bibnamefont {Ward}},\ and\
  \bibinfo {author} {\bibfnamefont {S.~M.}\ \bibnamefont {Schmon}},\ }\bibfield
   {title} {\bibinfo {title} {Investigating the impact of model
  misspecification in neural simulation-based inference},\ }\href@noop {}
  {\bibfield  {journal} {\bibinfo  {journal} {arXiv preprint arXiv:2209.01845}\
  } (\bibinfo {year} {2022})}\BibitemShut {NoStop}%
\bibitem [{\citenamefont {Rezende}\ and\ \citenamefont
  {Mohamed}(2015)}]{rezende2015variational}%
  \BibitemOpen
  \bibfield  {author} {\bibinfo {author} {\bibfnamefont {D.}~\bibnamefont
  {Rezende}}\ and\ \bibinfo {author} {\bibfnamefont {S.}~\bibnamefont
  {Mohamed}},\ }\bibfield  {title} {\bibinfo {title} {Variational inference
  with normalizing flows},\ }in\ \href@noop {} {\emph {\bibinfo {booktitle}
  {International Conference on Machine Learning}}}\ (\bibinfo {year} {2015})\
  pp.\ \bibinfo {pages} {1530--1538},\ \Eprint
  {https://arxiv.org/abs/1505.05770} {1505.05770 [stat.ML]} \BibitemShut
  {NoStop}%
\bibitem [{\citenamefont {Kingma}\ \emph {et~al.}(2016)\citenamefont {Kingma},
  \citenamefont {Salimans}, \citenamefont {Jozefowicz}, \citenamefont {Chen},
  \citenamefont {Sutskever},\ and\ \citenamefont
  {Welling}}]{kingma2016improved}%
  \BibitemOpen
  \bibfield  {author} {\bibinfo {author} {\bibfnamefont {D.~P.}\ \bibnamefont
  {Kingma}}, \bibinfo {author} {\bibfnamefont {T.}~\bibnamefont {Salimans}},
  \bibinfo {author} {\bibfnamefont {R.}~\bibnamefont {Jozefowicz}}, \bibinfo
  {author} {\bibfnamefont {X.}~\bibnamefont {Chen}}, \bibinfo {author}
  {\bibfnamefont {I.}~\bibnamefont {Sutskever}},\ and\ \bibinfo {author}
  {\bibfnamefont {M.}~\bibnamefont {Welling}},\ }\bibfield  {title} {\bibinfo
  {title} {Improved variational inference with inverse autoregressive flow},\
  }in\ \href@noop {} {\emph {\bibinfo {booktitle} {Advances in neural
  information processing systems}}}\ (\bibinfo {year} {2016})\ pp.\ \bibinfo
  {pages} {4743--4751},\ \Eprint {https://arxiv.org/abs/1606.04934}
  {arXiv:1606.04934 [cs.LG]} \BibitemShut {NoStop}%
\bibitem [{\citenamefont {Durkan}\ \emph {et~al.}(2019)\citenamefont {Durkan},
  \citenamefont {Bekasov}, \citenamefont {Murray},\ and\ \citenamefont
  {Papamakarios}}]{durkan2019neural}%
  \BibitemOpen
  \bibfield  {author} {\bibinfo {author} {\bibfnamefont {C.}~\bibnamefont
  {Durkan}}, \bibinfo {author} {\bibfnamefont {A.}~\bibnamefont {Bekasov}},
  \bibinfo {author} {\bibfnamefont {I.}~\bibnamefont {Murray}},\ and\ \bibinfo
  {author} {\bibfnamefont {G.}~\bibnamefont {Papamakarios}},\ }\bibfield
  {title} {\bibinfo {title} {Neural spline flows},\ }in\ \href@noop {} {\emph
  {\bibinfo {booktitle} {Advances in Neural Information Processing Systems}}}\
  (\bibinfo {year} {2019})\ pp.\ \bibinfo {pages} {7509--7520},\ \Eprint
  {https://arxiv.org/abs/1906.04032} {arXiv:1906.04032 [stat.ML]} \BibitemShut
  {NoStop}%
\bibitem [{\citenamefont {Papamakarios}\ \emph {et~al.}(2021)\citenamefont
  {Papamakarios}, \citenamefont {Nalisnick}, \citenamefont {Rezende},
  \citenamefont {Mohamed},\ and\ \citenamefont
  {Lakshminarayanan}}]{papamakarios2021normalizing}%
  \BibitemOpen
  \bibfield  {author} {\bibinfo {author} {\bibfnamefont {G.}~\bibnamefont
  {Papamakarios}}, \bibinfo {author} {\bibfnamefont {E.~T.}\ \bibnamefont
  {Nalisnick}}, \bibinfo {author} {\bibfnamefont {D.~J.}\ \bibnamefont
  {Rezende}}, \bibinfo {author} {\bibfnamefont {S.}~\bibnamefont {Mohamed}},\
  and\ \bibinfo {author} {\bibfnamefont {B.}~\bibnamefont {Lakshminarayanan}},\
  }\bibfield  {title} {\bibinfo {title} {Normalizing flows for probabilistic
  modeling and inference.},\ }\href@noop {} {\bibfield  {journal} {\bibinfo
  {journal} {J. Mach. Learn. Res.}\ }\textbf {\bibinfo {volume} {22}},\
  \bibinfo {pages} {1} (\bibinfo {year} {2021})}\BibitemShut {NoStop}%
\bibitem [{\citenamefont {Tokdar}\ and\ \citenamefont
  {Kass}(2010)}]{tokdar2010importance}%
  \BibitemOpen
  \bibfield  {author} {\bibinfo {author} {\bibfnamefont {S.~T.}\ \bibnamefont
  {Tokdar}}\ and\ \bibinfo {author} {\bibfnamefont {R.~E.}\ \bibnamefont
  {Kass}},\ }\bibfield  {title} {\bibinfo {title} {Importance sampling: a
  review},\ }\href@noop {} {\bibfield  {journal} {\bibinfo  {journal} {Wiley
  Interdisciplinary Reviews: Computational Statistics}\ }\textbf {\bibinfo
  {volume} {2}},\ \bibinfo {pages} {54} (\bibinfo {year} {2010})}\BibitemShut
  {NoStop}%
\bibitem [{\citenamefont {Veitch}\ \emph {et~al.}(2015)\citenamefont {Veitch},
  \citenamefont {Raymond}, \citenamefont {Farr}, \citenamefont {Farr},
  \citenamefont {Graff}, \citenamefont {Vitale} \emph
  {et~al.}}]{Veitch:2014wba}%
  \BibitemOpen
  \bibfield  {author} {\bibinfo {author} {\bibfnamefont {J.}~\bibnamefont
  {Veitch}}, \bibinfo {author} {\bibfnamefont {V.}~\bibnamefont {Raymond}},
  \bibinfo {author} {\bibfnamefont {B.}~\bibnamefont {Farr}}, \bibinfo {author}
  {\bibfnamefont {W.}~\bibnamefont {Farr}}, \bibinfo {author} {\bibfnamefont
  {P.}~\bibnamefont {Graff}}, \bibinfo {author} {\bibfnamefont
  {S.}~\bibnamefont {Vitale}}, \emph {et~al.},\ }\bibfield  {title} {\bibinfo
  {title} {{Parameter estimation for compact binaries with ground-based
  gravitational-wave observations using the LALInference software library}},\
  }\href {https://doi.org/10.1103/PhysRevD.91.042003} {\bibfield  {journal}
  {\bibinfo  {journal} {Phys. Rev.}\ }\textbf {\bibinfo {volume} {D91}},\
  \bibinfo {pages} {042003} (\bibinfo {year} {2015})},\ \Eprint
  {https://arxiv.org/abs/1409.7215} {arXiv:1409.7215 [gr-qc]} \BibitemShut
  {NoStop}%
\bibitem [{\citenamefont {Ashton}\ \emph {et~al.}(2019)\citenamefont {Ashton}
  \emph {et~al.}}]{Ashton:2018jfp}%
  \BibitemOpen
  \bibfield  {author} {\bibinfo {author} {\bibfnamefont {G.}~\bibnamefont
  {Ashton}} \emph {et~al.},\ }\bibfield  {title} {\bibinfo {title} {{BILBY: A
  user-friendly Bayesian inference library for gravitational-wave astronomy}},\
  }\href {https://doi.org/10.3847/1538-4365/ab06fc} {\bibfield  {journal}
  {\bibinfo  {journal} {Astrophys. J. Suppl.}\ }\textbf {\bibinfo {volume}
  {241}},\ \bibinfo {pages} {27} (\bibinfo {year} {2019})},\ \Eprint
  {https://arxiv.org/abs/1811.02042} {arXiv:1811.02042 [astro-ph.IM]}
  \BibitemShut {NoStop}%
\bibitem [{\citenamefont {Romero-Shaw}\ \emph {et~al.}(2020)\citenamefont
  {Romero-Shaw} \emph {et~al.}}]{Romero-Shaw:2020owr}%
  \BibitemOpen
  \bibfield  {author} {\bibinfo {author} {\bibfnamefont {I.~M.}\ \bibnamefont
  {Romero-Shaw}} \emph {et~al.},\ }\bibfield  {title} {\bibinfo {title}
  {{Bayesian inference for compact binary coalescences with bilby: validation
  and application to the first LIGO\textendash{}Virgo gravitational-wave
  transient catalogue}},\ }\href {https://doi.org/10.1093/mnras/staa2850}
  {\bibfield  {journal} {\bibinfo  {journal} {Mon. Not. Roy. Astron. Soc.}\
  }\textbf {\bibinfo {volume} {499}},\ \bibinfo {pages} {3295} (\bibinfo {year}
  {2020})},\ \Eprint {https://arxiv.org/abs/2006.00714} {arXiv:2006.00714
  [astro-ph.IM]} \BibitemShut {NoStop}%
\bibitem [{\citenamefont {Speagle}(2020)}]{Speagle_2020}%
  \BibitemOpen
  \bibfield  {author} {\bibinfo {author} {\bibfnamefont {J.~S.}\ \bibnamefont
  {Speagle}},\ }\bibfield  {title} {\bibinfo {title} {dynesty: a dynamic nested
  sampling package for estimating bayesian posteriors and evidences},\ }\href
  {https://doi.org/10.1093/mnras/staa278} {\bibfield  {journal} {\bibinfo
  {journal} {Monthly Notices of the Royal Astronomical Society}\ }\textbf
  {\bibinfo {volume} {493}},\ \bibinfo {pages} {3132–3158} (\bibinfo {year}
  {2020})},\ \Eprint {https://arxiv.org/abs/1904.02180} {arXiv:1904.02180
  [astro-ph.IM]} \BibitemShut {NoStop}%
\bibitem [{\citenamefont {Abbott}\ \emph
  {et~al.}(2021{\natexlab{e}})\citenamefont {Abbott} \emph
  {et~al.}}]{LIGOScientific:2020ibl}%
  \BibitemOpen
  \bibfield  {author} {\bibinfo {author} {\bibfnamefont {R.}~\bibnamefont
  {Abbott}} \emph {et~al.} (\bibinfo {collaboration} {LIGO Scientific,
  Virgo}),\ }\bibfield  {title} {\bibinfo {title} {{GWTC-2: Compact Binary
  Coalescences Observed by LIGO and Virgo During the First Half of the Third
  Observing Run}},\ }\href {https://doi.org/10.1103/PhysRevX.11.021053}
  {\bibfield  {journal} {\bibinfo  {journal} {Phys. Rev. X}\ }\textbf {\bibinfo
  {volume} {11}},\ \bibinfo {pages} {021053} (\bibinfo {year}
  {2021}{\natexlab{e}})},\ \Eprint {https://arxiv.org/abs/2010.14527}
  {arXiv:2010.14527 [gr-qc]} \BibitemShut {NoStop}%
\bibitem [{\citenamefont {Pratten}\ \emph {et~al.}(2021)\citenamefont {Pratten}
  \emph {et~al.}}]{Pratten:2020ceb}%
  \BibitemOpen
  \bibfield  {author} {\bibinfo {author} {\bibfnamefont {G.}~\bibnamefont
  {Pratten}} \emph {et~al.},\ }\bibfield  {title} {\bibinfo {title}
  {{Computationally efficient models for the dominant and subdominant harmonic
  modes of precessing binary black holes}},\ }\href
  {https://doi.org/10.1103/PhysRevD.103.104056} {\bibfield  {journal} {\bibinfo
   {journal} {Phys. Rev. D}\ }\textbf {\bibinfo {volume} {103}},\ \bibinfo
  {pages} {104056} (\bibinfo {year} {2021})},\ \Eprint
  {https://arxiv.org/abs/2004.06503} {arXiv:2004.06503 [gr-qc]} \BibitemShut
  {NoStop}%
\bibitem [{\citenamefont {Ossokine}\ \emph {et~al.}(2020)\citenamefont
  {Ossokine} \emph {et~al.}}]{Ossokine:2020kjp}%
  \BibitemOpen
  \bibfield  {author} {\bibinfo {author} {\bibfnamefont {S.}~\bibnamefont
  {Ossokine}} \emph {et~al.},\ }\bibfield  {title} {\bibinfo {title}
  {{Multipolar Effective-One-Body Waveforms for Precessing Binary Black Holes:
  Construction and Validation}},\ }\href
  {https://doi.org/10.1103/PhysRevD.102.044055} {\bibfield  {journal} {\bibinfo
   {journal} {Phys. Rev. D}\ }\textbf {\bibinfo {volume} {102}},\ \bibinfo
  {pages} {044055} (\bibinfo {year} {2020})},\ \Eprint
  {https://arxiv.org/abs/2004.09442} {arXiv:2004.09442 [gr-qc]} \BibitemShut
  {NoStop}%
\bibitem [{\citenamefont {Cuoco}\ \emph {et~al.}(2020)\citenamefont {Cuoco},
  \citenamefont {Powell}, \citenamefont {Cavagli{\`a}}, \citenamefont {Ackley},
  \citenamefont {Bejger}, \citenamefont {Chatterjee}, \citenamefont {Coughlin},
  \citenamefont {Coughlin}, \citenamefont {Easter}, \citenamefont {Essick}
  \emph {et~al.}}]{Cuoco:2020ogp}%
  \BibitemOpen
  \bibfield  {author} {\bibinfo {author} {\bibfnamefont {E.}~\bibnamefont
  {Cuoco}}, \bibinfo {author} {\bibfnamefont {J.}~\bibnamefont {Powell}},
  \bibinfo {author} {\bibfnamefont {M.}~\bibnamefont {Cavagli{\`a}}}, \bibinfo
  {author} {\bibfnamefont {K.}~\bibnamefont {Ackley}}, \bibinfo {author}
  {\bibfnamefont {M.}~\bibnamefont {Bejger}}, \bibinfo {author} {\bibfnamefont
  {C.}~\bibnamefont {Chatterjee}}, \bibinfo {author} {\bibfnamefont
  {M.}~\bibnamefont {Coughlin}}, \bibinfo {author} {\bibfnamefont
  {S.}~\bibnamefont {Coughlin}}, \bibinfo {author} {\bibfnamefont
  {P.}~\bibnamefont {Easter}}, \bibinfo {author} {\bibfnamefont
  {R.}~\bibnamefont {Essick}}, \emph {et~al.},\ }\bibfield  {title} {\bibinfo
  {title} {Enhancing gravitational-wave science with machine learning},\ }\href
  {https://doi.org/10.1088/2632-2153/abb93a} {\bibfield  {journal} {\bibinfo
  {journal} {Machine Learning: Science and Technology}\ }\textbf {\bibinfo
  {volume} {2}},\ \bibinfo {pages} {011002} (\bibinfo {year} {2020})},\ \Eprint
  {https://arxiv.org/abs/2005.03745} {arXiv:2005.03745 [astro-ph.HE]}
  \BibitemShut {NoStop}%
\bibitem [{\citenamefont {Gabbard}\ \emph {et~al.}(2022)\citenamefont
  {Gabbard}, \citenamefont {Messenger}, \citenamefont {Heng}, \citenamefont
  {Tonolini},\ and\ \citenamefont {Murray-Smith}}]{Gabbard:2019rde}%
  \BibitemOpen
  \bibfield  {author} {\bibinfo {author} {\bibfnamefont {H.}~\bibnamefont
  {Gabbard}}, \bibinfo {author} {\bibfnamefont {C.}~\bibnamefont {Messenger}},
  \bibinfo {author} {\bibfnamefont {I.~S.}\ \bibnamefont {Heng}}, \bibinfo
  {author} {\bibfnamefont {F.}~\bibnamefont {Tonolini}},\ and\ \bibinfo
  {author} {\bibfnamefont {R.}~\bibnamefont {Murray-Smith}},\ }\bibfield
  {title} {\bibinfo {title} {{Bayesian parameter estimation using conditional
  variational autoencoders for gravitational-wave astronomy}},\ }\href
  {https://doi.org/10.1038/s41567-021-01425-7} {\bibfield  {journal} {\bibinfo
  {journal} {Nature Phys.}\ }\textbf {\bibinfo {volume} {18}},\ \bibinfo
  {pages} {112} (\bibinfo {year} {2022})},\ \Eprint
  {https://arxiv.org/abs/1909.06296} {arXiv:1909.06296 [astro-ph.IM]}
  \BibitemShut {NoStop}%
\bibitem [{\citenamefont {Green}\ \emph {et~al.}(2020)\citenamefont {Green},
  \citenamefont {Simpson},\ and\ \citenamefont {Gair}}]{Green:2020hst}%
  \BibitemOpen
  \bibfield  {author} {\bibinfo {author} {\bibfnamefont {S.~R.}\ \bibnamefont
  {Green}}, \bibinfo {author} {\bibfnamefont {C.}~\bibnamefont {Simpson}},\
  and\ \bibinfo {author} {\bibfnamefont {J.}~\bibnamefont {Gair}},\ }\bibfield
  {title} {\bibinfo {title} {{Gravitational-wave parameter estimation with
  autoregressive neural network flows}},\ }\href
  {https://doi.org/10.1103/PhysRevD.102.104057} {\bibfield  {journal} {\bibinfo
   {journal} {Phys. Rev. D}\ }\textbf {\bibinfo {volume} {102}},\ \bibinfo
  {pages} {104057} (\bibinfo {year} {2020})},\ \Eprint
  {https://arxiv.org/abs/2002.07656} {arXiv:2002.07656 [astro-ph.IM]}
  \BibitemShut {NoStop}%
\bibitem [{\citenamefont {Delaunoy}\ \emph {et~al.}(2020)\citenamefont
  {Delaunoy}, \citenamefont {Wehenkel}, \citenamefont {Hinderer}, \citenamefont
  {Nissanke}, \citenamefont {Weniger}, \citenamefont {Williamson},\ and\
  \citenamefont {Louppe}}]{Delaunoy:2020zcu}%
  \BibitemOpen
  \bibfield  {author} {\bibinfo {author} {\bibfnamefont {A.}~\bibnamefont
  {Delaunoy}}, \bibinfo {author} {\bibfnamefont {A.}~\bibnamefont {Wehenkel}},
  \bibinfo {author} {\bibfnamefont {T.}~\bibnamefont {Hinderer}}, \bibinfo
  {author} {\bibfnamefont {S.}~\bibnamefont {Nissanke}}, \bibinfo {author}
  {\bibfnamefont {C.}~\bibnamefont {Weniger}}, \bibinfo {author} {\bibfnamefont
  {A.~R.}\ \bibnamefont {Williamson}},\ and\ \bibinfo {author} {\bibfnamefont
  {G.}~\bibnamefont {Louppe}},\ }\bibfield  {title} {\bibinfo {title}
  {{Lightning-Fast Gravitational Wave Parameter Inference through Neural
  Amortization}},\ }in\ \href@noop {} {\emph {\bibinfo {booktitle} {Third
  Workshop on Machine Learning and the Physical Sciences}}}\ (\bibinfo {year}
  {2020})\ \Eprint {https://arxiv.org/abs/2010.12931} {arXiv:2010.12931
  [astro-ph.IM]} \BibitemShut {NoStop}%
\bibitem [{\citenamefont {Green}\ and\ \citenamefont
  {Gair}(2021)}]{Green:2020dnx}%
  \BibitemOpen
  \bibfield  {author} {\bibinfo {author} {\bibfnamefont {S.~R.}\ \bibnamefont
  {Green}}\ and\ \bibinfo {author} {\bibfnamefont {J.}~\bibnamefont {Gair}},\
  }\bibfield  {title} {\bibinfo {title} {{Complete parameter inference for
  GW150914 using deep learning}},\ }\href
  {https://doi.org/10.1088/2632-2153/abfaed} {\bibfield  {journal} {\bibinfo
  {journal} {Mach. Learn. Sci. Tech.}\ }\textbf {\bibinfo {volume} {2}},\
  \bibinfo {pages} {03LT01} (\bibinfo {year} {2021})},\ \Eprint
  {https://arxiv.org/abs/2008.03312} {arXiv:2008.03312 [astro-ph.IM]}
  \BibitemShut {NoStop}%
\bibitem [{\citenamefont {Chatterjee}\ \emph {et~al.}(2022)\citenamefont
  {Chatterjee}, \citenamefont {Wen}, \citenamefont {Beveridge}, \citenamefont
  {Diakogiannis},\ and\ \citenamefont {Vinsen}}]{Chatterjee:2022ggk}%
  \BibitemOpen
  \bibfield  {author} {\bibinfo {author} {\bibfnamefont {C.}~\bibnamefont
  {Chatterjee}}, \bibinfo {author} {\bibfnamefont {L.}~\bibnamefont {Wen}},
  \bibinfo {author} {\bibfnamefont {D.}~\bibnamefont {Beveridge}}, \bibinfo
  {author} {\bibfnamefont {F.}~\bibnamefont {Diakogiannis}},\ and\ \bibinfo
  {author} {\bibfnamefont {K.}~\bibnamefont {Vinsen}},\ }\bibfield  {title}
  {\bibinfo {title} {{Rapid localization of gravitational wave sources from
  compact binary coalescences using deep learning}},\ }\href@noop {} {\
  (\bibinfo {year} {2022})},\ \Eprint {https://arxiv.org/abs/2207.14522}
  {arXiv:2207.14522 [gr-qc]} \BibitemShut {NoStop}%
\bibitem [{\citenamefont {Williams}\ \emph {et~al.}(2021)\citenamefont
  {Williams}, \citenamefont {Veitch},\ and\ \citenamefont
  {Messenger}}]{Williams:2021qyt}%
  \BibitemOpen
  \bibfield  {author} {\bibinfo {author} {\bibfnamefont {M.~J.}\ \bibnamefont
  {Williams}}, \bibinfo {author} {\bibfnamefont {J.}~\bibnamefont {Veitch}},\
  and\ \bibinfo {author} {\bibfnamefont {C.}~\bibnamefont {Messenger}},\
  }\bibfield  {title} {\bibinfo {title} {{Nested sampling with normalizing
  flows for gravitational-wave inference}},\ }\href
  {https://doi.org/10.1103/PhysRevD.103.103006} {\bibfield  {journal} {\bibinfo
   {journal} {Phys. Rev. D}\ }\textbf {\bibinfo {volume} {103}},\ \bibinfo
  {pages} {103006} (\bibinfo {year} {2021})},\ \Eprint
  {https://arxiv.org/abs/2102.11056} {arXiv:2102.11056 [gr-qc]} \BibitemShut
  {NoStop}%
\bibitem [{\citenamefont {Paige}\ and\ \citenamefont
  {Wood}(2016)}]{paige2016inference}%
  \BibitemOpen
  \bibfield  {author} {\bibinfo {author} {\bibfnamefont {B.}~\bibnamefont
  {Paige}}\ and\ \bibinfo {author} {\bibfnamefont {F.}~\bibnamefont {Wood}},\
  }\bibfield  {title} {\bibinfo {title} {Inference networks for sequential
  monte carlo in graphical models},\ }in\ \href@noop {} {\emph {\bibinfo
  {booktitle} {International Conference on Machine Learning}}}\ (\bibinfo
  {organization} {PMLR},\ \bibinfo {year} {2016})\ pp.\ \bibinfo {pages}
  {3040--3049},\ \Eprint {https://arxiv.org/abs/1602.06701} {arXiv:1602.06701
  [stat.ML]} \BibitemShut {NoStop}%
\bibitem [{\citenamefont {M{\"u}ller}\ \emph {et~al.}(2019)\citenamefont
  {M{\"u}ller}, \citenamefont {McWilliams}, \citenamefont {Rousselle},
  \citenamefont {Gross},\ and\ \citenamefont {Nov{\'a}k}}]{muller2019neural}%
  \BibitemOpen
  \bibfield  {author} {\bibinfo {author} {\bibfnamefont {T.}~\bibnamefont
  {M{\"u}ller}}, \bibinfo {author} {\bibfnamefont {B.}~\bibnamefont
  {McWilliams}}, \bibinfo {author} {\bibfnamefont {F.}~\bibnamefont
  {Rousselle}}, \bibinfo {author} {\bibfnamefont {M.}~\bibnamefont {Gross}},\
  and\ \bibinfo {author} {\bibfnamefont {J.}~\bibnamefont {Nov{\'a}k}},\
  }\bibfield  {title} {\bibinfo {title} {Neural importance sampling},\
  }\href@noop {} {\bibfield  {journal} {\bibinfo  {journal} {ACM Transactions
  on Graphics (TOG)}\ }\textbf {\bibinfo {volume} {38}},\ \bibinfo {pages} {1}
  (\bibinfo {year} {2019})}\BibitemShut {NoStop}%
\bibitem [{\citenamefont {No{\'e}}\ \emph {et~al.}(2019)\citenamefont
  {No{\'e}}, \citenamefont {Olsson}, \citenamefont {K{\"o}hler},\ and\
  \citenamefont {Wu}}]{noe2019boltzmann}%
  \BibitemOpen
  \bibfield  {author} {\bibinfo {author} {\bibfnamefont {F.}~\bibnamefont
  {No{\'e}}}, \bibinfo {author} {\bibfnamefont {S.}~\bibnamefont {Olsson}},
  \bibinfo {author} {\bibfnamefont {J.}~\bibnamefont {K{\"o}hler}},\ and\
  \bibinfo {author} {\bibfnamefont {H.}~\bibnamefont {Wu}},\ }\bibfield
  {title} {\bibinfo {title} {Boltzmann generators: Sampling equilibrium states
  of many-body systems with deep learning},\ }\href@noop {} {\bibfield
  {journal} {\bibinfo  {journal} {Science}\ }\textbf {\bibinfo {volume}
  {365}},\ \bibinfo {pages} {eaaw1147} (\bibinfo {year} {2019})}\BibitemShut
  {NoStop}%
\bibitem [{\citenamefont {Albergo}\ \emph {et~al.}(2019)\citenamefont
  {Albergo}, \citenamefont {Kanwar},\ and\ \citenamefont
  {Shanahan}}]{Albergo:2019eim}%
  \BibitemOpen
  \bibfield  {author} {\bibinfo {author} {\bibfnamefont {M.~S.}\ \bibnamefont
  {Albergo}}, \bibinfo {author} {\bibfnamefont {G.}~\bibnamefont {Kanwar}},\
  and\ \bibinfo {author} {\bibfnamefont {P.~E.}\ \bibnamefont {Shanahan}},\
  }\bibfield  {title} {\bibinfo {title} {{Flow-based generative models for
  Markov chain Monte Carlo in lattice field theory}},\ }\href
  {https://doi.org/10.1103/PhysRevD.100.034515} {\bibfield  {journal} {\bibinfo
   {journal} {Phys. Rev. D}\ }\textbf {\bibinfo {volume} {100}},\ \bibinfo
  {pages} {034515} (\bibinfo {year} {2019})},\ \Eprint
  {https://arxiv.org/abs/1904.12072} {arXiv:1904.12072 [hep-lat]} \BibitemShut
  {NoStop}%
\bibitem [{\citenamefont {Kanwar}\ \emph {et~al.}(2020)\citenamefont {Kanwar},
  \citenamefont {Albergo}, \citenamefont {Boyda}, \citenamefont {Cranmer},
  \citenamefont {Hackett}, \citenamefont {Racani\`ere}, \citenamefont
  {Rezende},\ and\ \citenamefont {Shanahan}}]{Kanwar:2020xzo}%
  \BibitemOpen
  \bibfield  {author} {\bibinfo {author} {\bibfnamefont {G.}~\bibnamefont
  {Kanwar}}, \bibinfo {author} {\bibfnamefont {M.~S.}\ \bibnamefont {Albergo}},
  \bibinfo {author} {\bibfnamefont {D.}~\bibnamefont {Boyda}}, \bibinfo
  {author} {\bibfnamefont {K.}~\bibnamefont {Cranmer}}, \bibinfo {author}
  {\bibfnamefont {D.~C.}\ \bibnamefont {Hackett}}, \bibinfo {author}
  {\bibfnamefont {S.}~\bibnamefont {Racani\`ere}}, \bibinfo {author}
  {\bibfnamefont {D.~J.}\ \bibnamefont {Rezende}},\ and\ \bibinfo {author}
  {\bibfnamefont {P.~E.}\ \bibnamefont {Shanahan}},\ }\bibfield  {title}
  {\bibinfo {title} {{Equivariant flow-based sampling for lattice gauge
  theory}},\ }\href {https://doi.org/10.1103/PhysRevLett.125.121601} {\bibfield
   {journal} {\bibinfo  {journal} {Phys. Rev. Lett.}\ }\textbf {\bibinfo
  {volume} {125}},\ \bibinfo {pages} {121601} (\bibinfo {year} {2020})},\
  \Eprint {https://arxiv.org/abs/2003.06413} {arXiv:2003.06413 [hep-lat]}
  \BibitemShut {NoStop}%
\bibitem [{\citenamefont {Sun}\ \emph {et~al.}(2022)\citenamefont {Sun},
  \citenamefont {Bouman}, \citenamefont {Tiede}, \citenamefont {Wang},
  \citenamefont {Blunt},\ and\ \citenamefont {Mawet}}]{sun2022alpha}%
  \BibitemOpen
  \bibfield  {author} {\bibinfo {author} {\bibfnamefont {H.}~\bibnamefont
  {Sun}}, \bibinfo {author} {\bibfnamefont {K.~L.}\ \bibnamefont {Bouman}},
  \bibinfo {author} {\bibfnamefont {P.}~\bibnamefont {Tiede}}, \bibinfo
  {author} {\bibfnamefont {J.~J.}\ \bibnamefont {Wang}}, \bibinfo {author}
  {\bibfnamefont {S.}~\bibnamefont {Blunt}},\ and\ \bibinfo {author}
  {\bibfnamefont {D.}~\bibnamefont {Mawet}},\ }\bibfield  {title} {\bibinfo
  {title} {$\alpha$-deep probabilistic inference ($\alpha$-dpi): Efficient
  uncertainty quantification from exoplanet astrometry to black hole feature
  extraction},\ }\href {https://doi.org/10.3847/1538-4357/ac6be9} {\bibfield
  {journal} {\bibinfo  {journal} {The Astrophysical Journal}\ }\textbf
  {\bibinfo {volume} {932}},\ \bibinfo {pages} {99} (\bibinfo {year}
  {2022})}\BibitemShut {NoStop}%
\bibitem [{\citenamefont {Kolmus}\ \emph {et~al.}(2022)\citenamefont {Kolmus},
  \citenamefont {Baltus}, \citenamefont {Janquart}, \citenamefont {van
  Laarhoven}, \citenamefont {Caudill},\ and\ \citenamefont
  {Heskes}}]{Kolmus:2021buf}%
  \BibitemOpen
  \bibfield  {author} {\bibinfo {author} {\bibfnamefont {A.}~\bibnamefont
  {Kolmus}}, \bibinfo {author} {\bibfnamefont {G.}~\bibnamefont {Baltus}},
  \bibinfo {author} {\bibfnamefont {J.}~\bibnamefont {Janquart}}, \bibinfo
  {author} {\bibfnamefont {T.}~\bibnamefont {van Laarhoven}}, \bibinfo {author}
  {\bibfnamefont {S.}~\bibnamefont {Caudill}},\ and\ \bibinfo {author}
  {\bibfnamefont {T.}~\bibnamefont {Heskes}},\ }\bibfield  {title} {\bibinfo
  {title} {{Fast sky localization of gravitational waves using deep learning
  seeded importance sampling}},\ }\href
  {https://doi.org/10.1103/PhysRevD.106.023032} {\bibfield  {journal} {\bibinfo
   {journal} {Phys. Rev. D}\ }\textbf {\bibinfo {volume} {106}},\ \bibinfo
  {pages} {023032} (\bibinfo {year} {2022})},\ \Eprint
  {https://arxiv.org/abs/2111.00833} {arXiv:2111.00833 [gr-qc]} \BibitemShut
  {NoStop}%
\bibitem [{\citenamefont {Papamakarios}\ and\ \citenamefont
  {Murray}(2016)}]{papamakarios2016fast}%
  \BibitemOpen
  \bibfield  {author} {\bibinfo {author} {\bibfnamefont {G.}~\bibnamefont
  {Papamakarios}}\ and\ \bibinfo {author} {\bibfnamefont {I.}~\bibnamefont
  {Murray}},\ }\bibfield  {title} {\bibinfo {title} {Fast $\varepsilon$-free
  inference of simulation models with bayesian conditional density
  estimation},\ }in\ \href@noop {} {\emph {\bibinfo {booktitle} {Advances in
  neural information processing systems}}}\ (\bibinfo {year} {2016})\ \Eprint
  {https://arxiv.org/abs/1605.06376} {arXiv:1605.06376 [stat.ML]} \BibitemShut
  {NoStop}%
\bibitem [{\citenamefont {Kong}(1992)}]{kong1992note}%
  \BibitemOpen
  \bibfield  {author} {\bibinfo {author} {\bibfnamefont {A.}~\bibnamefont
  {Kong}},\ }\bibfield  {title} {\bibinfo {title} {A note on importance
  sampling using standardized weights},\ }\href@noop {} {\bibfield  {journal}
  {\bibinfo  {journal} {University of Chicago, Dept. of Statistics, Tech. Rep}\
  }\textbf {\bibinfo {volume} {348}} (\bibinfo {year} {1992})}\BibitemShut
  {NoStop}%
\bibitem [{\citenamefont {Owen}(2013)}]{mcbook}%
  \BibitemOpen
  \bibfield  {author} {\bibinfo {author} {\bibfnamefont {A.~B.}\ \bibnamefont
  {Owen}},\ }\href@noop {} {\emph {\bibinfo {title} {Monte Carlo theory,
  methods and examples}}}\ (\bibinfo {year} {2013})\BibitemShut {NoStop}%
\bibitem [{\citenamefont {Jordan}\ \emph {et~al.}(1999)\citenamefont {Jordan},
  \citenamefont {Ghahramani}, \citenamefont {Jaakkola},\ and\ \citenamefont
  {Saul}}]{jordan1999introduction}%
  \BibitemOpen
  \bibfield  {author} {\bibinfo {author} {\bibfnamefont {M.~I.}\ \bibnamefont
  {Jordan}}, \bibinfo {author} {\bibfnamefont {Z.}~\bibnamefont {Ghahramani}},
  \bibinfo {author} {\bibfnamefont {T.~S.}\ \bibnamefont {Jaakkola}},\ and\
  \bibinfo {author} {\bibfnamefont {L.~K.}\ \bibnamefont {Saul}},\ }\bibfield
  {title} {\bibinfo {title} {An introduction to variational methods for
  graphical models},\ }\href@noop {} {\bibfield  {journal} {\bibinfo  {journal}
  {Machine learning}\ }\textbf {\bibinfo {volume} {37}},\ \bibinfo {pages}
  {183} (\bibinfo {year} {1999})}\BibitemShut {NoStop}%
\bibitem [{\citenamefont {Wainwright}\ \emph {et~al.}(2008)\citenamefont
  {Wainwright}, \citenamefont {Jordan} \emph
  {et~al.}}]{wainwright2008graphical}%
  \BibitemOpen
  \bibfield  {author} {\bibinfo {author} {\bibfnamefont {M.~J.}\ \bibnamefont
  {Wainwright}}, \bibinfo {author} {\bibfnamefont {M.~I.}\ \bibnamefont
  {Jordan}}, \emph {et~al.},\ }\bibfield  {title} {\bibinfo {title} {Graphical
  models, exponential families, and variational inference},\ }\href@noop {}
  {\bibfield  {journal} {\bibinfo  {journal} {Foundations and
  Trends{\textregistered} in Machine Learning}\ }\textbf {\bibinfo {volume}
  {1}},\ \bibinfo {pages} {1} (\bibinfo {year} {2008})}\BibitemShut {NoStop}%
\bibitem [{\citenamefont {Papamakarios}\ \emph {et~al.}(2017)\citenamefont
  {Papamakarios}, \citenamefont {Pavlakou},\ and\ \citenamefont
  {Murray}}]{papamakarios2017masked}%
  \BibitemOpen
  \bibfield  {author} {\bibinfo {author} {\bibfnamefont {G.}~\bibnamefont
  {Papamakarios}}, \bibinfo {author} {\bibfnamefont {T.}~\bibnamefont
  {Pavlakou}},\ and\ \bibinfo {author} {\bibfnamefont {I.}~\bibnamefont
  {Murray}},\ }\bibfield  {title} {\bibinfo {title} {Masked autoregressive flow
  for density estimation},\ }in\ \href@noop {} {\emph {\bibinfo {booktitle}
  {Advances in Neural Information Processing Systems}}}\ (\bibinfo {year}
  {2017})\ pp.\ \bibinfo {pages} {2338--2347},\ \Eprint
  {https://arxiv.org/abs/1705.07057} {arXiv:1705.07057 [stat.ML]} \BibitemShut
  {NoStop}%
\bibitem [{\citenamefont {Aasi}\ \emph {et~al.}(2015)\citenamefont {Aasi} \emph
  {et~al.}}]{LIGOScientific:2014pky}%
  \BibitemOpen
  \bibfield  {author} {\bibinfo {author} {\bibfnamefont {J.}~\bibnamefont
  {Aasi}} \emph {et~al.} (\bibinfo {collaboration} {LIGO Scientific}),\
  }\bibfield  {title} {\bibinfo {title} {{Advanced LIGO}},\ }\href
  {https://doi.org/10.1088/0264-9381/32/7/074001} {\bibfield  {journal}
  {\bibinfo  {journal} {Class. Quant. Grav.}\ }\textbf {\bibinfo {volume}
  {32}},\ \bibinfo {pages} {074001} (\bibinfo {year} {2015})},\ \Eprint
  {https://arxiv.org/abs/1411.4547} {arXiv:1411.4547 [gr-qc]} \BibitemShut
  {NoStop}%
\bibitem [{\citenamefont {Acernese}\ \emph {et~al.}(2015)\citenamefont
  {Acernese} \emph {et~al.}}]{VIRGO:2014yos}%
  \BibitemOpen
  \bibfield  {author} {\bibinfo {author} {\bibfnamefont {F.}~\bibnamefont
  {Acernese}} \emph {et~al.} (\bibinfo {collaboration} {VIRGO}),\ }\bibfield
  {title} {\bibinfo {title} {{Advanced Virgo: a second-generation
  interferometric gravitational wave detector}},\ }\href
  {https://doi.org/10.1088/0264-9381/32/2/024001} {\bibfield  {journal}
  {\bibinfo  {journal} {Class. Quant. Grav.}\ }\textbf {\bibinfo {volume}
  {32}},\ \bibinfo {pages} {024001} (\bibinfo {year} {2015})},\ \Eprint
  {https://arxiv.org/abs/1408.3978} {arXiv:1408.3978 [gr-qc]} \BibitemShut
  {NoStop}%
\bibitem [{\citenamefont {Veitch}\ and\ \citenamefont
  {Del~Pozzo}(2013)}]{veitch2013analytic}%
  \BibitemOpen
  \bibfield  {author} {\bibinfo {author} {\bibfnamefont {J.}~\bibnamefont
  {Veitch}}\ and\ \bibinfo {author} {\bibfnamefont {W.}~\bibnamefont
  {Del~Pozzo}},\ }\bibfield  {title} {\bibinfo {title} {Analytic
  marginalisation of phase parameter},\ }\href@noop {} {\bibfield  {journal}
  {\bibinfo  {journal} {URL: https://dcc. ligo. org/LIGO-T1300326/public}\ }
  (\bibinfo {year} {2013})}\BibitemShut {NoStop}%
\bibitem [{\citenamefont {Thrane}\ and\ \citenamefont
  {Talbot}(2019)}]{Thrane:2018qnx}%
  \BibitemOpen
  \bibfield  {author} {\bibinfo {author} {\bibfnamefont {E.}~\bibnamefont
  {Thrane}}\ and\ \bibinfo {author} {\bibfnamefont {C.}~\bibnamefont
  {Talbot}},\ }\bibfield  {title} {\bibinfo {title} {{An introduction to
  Bayesian inference in gravitational-wave astronomy: parameter estimation,
  model selection, and hierarchical models}},\ }\href
  {https://doi.org/10.1017/pasa.2019.2} {\bibfield  {journal} {\bibinfo
  {journal} {Publ. Astron. Soc. Austral.}\ }\textbf {\bibinfo {volume} {36}},\
  \bibinfo {pages} {e010} (\bibinfo {year} {2019})},\ \bibinfo {note}
  {[Erratum: Publ.Astron.Soc.Austral. 37, e036 (2020)]},\ \Eprint
  {https://arxiv.org/abs/1809.02293} {arXiv:1809.02293 [astro-ph.IM]}
  \BibitemShut {NoStop}%
\bibitem [{\citenamefont {Dax}\ \emph {et~al.}(2022)\citenamefont {Dax},
  \citenamefont {Green}, \citenamefont {Gair}, \citenamefont {Deistler},
  \citenamefont {Sch\"olkopf},\ and\ \citenamefont {Macke}}]{Dax:2021myb}%
  \BibitemOpen
  \bibfield  {author} {\bibinfo {author} {\bibfnamefont {M.}~\bibnamefont
  {Dax}}, \bibinfo {author} {\bibfnamefont {S.~R.}\ \bibnamefont {Green}},
  \bibinfo {author} {\bibfnamefont {J.}~\bibnamefont {Gair}}, \bibinfo {author}
  {\bibfnamefont {M.}~\bibnamefont {Deistler}}, \bibinfo {author}
  {\bibfnamefont {B.}~\bibnamefont {Sch\"olkopf}},\ and\ \bibinfo {author}
  {\bibfnamefont {J.~H.}\ \bibnamefont {Macke}},\ }\bibfield  {title} {\bibinfo
  {title} {{Group equivariant neural posterior estimation}},\ }in\ \href@noop
  {} {\emph {\bibinfo {booktitle} {International Conference on Learning
  Representations}}}\ (\bibinfo {year} {2022})\ \Eprint
  {https://arxiv.org/abs/2111.13139} {arXiv:2111.13139 [cs.LG]} \BibitemShut
  {NoStop}%
\bibitem [{\citenamefont {Pankow}\ \emph {et~al.}(2015)\citenamefont {Pankow},
  \citenamefont {Brady}, \citenamefont {Ochsner},\ and\ \citenamefont
  {O'Shaughnessy}}]{Pankow:2015cra}%
  \BibitemOpen
  \bibfield  {author} {\bibinfo {author} {\bibfnamefont {C.}~\bibnamefont
  {Pankow}}, \bibinfo {author} {\bibfnamefont {P.}~\bibnamefont {Brady}},
  \bibinfo {author} {\bibfnamefont {E.}~\bibnamefont {Ochsner}},\ and\ \bibinfo
  {author} {\bibfnamefont {R.}~\bibnamefont {O'Shaughnessy}},\ }\bibfield
  {title} {\bibinfo {title} {{Novel scheme for rapid parallel parameter
  estimation of gravitational waves from compact binary coalescences}},\ }\href
  {https://doi.org/10.1103/PhysRevD.92.023002} {\bibfield  {journal} {\bibinfo
  {journal} {Phys. Rev. D}\ }\textbf {\bibinfo {volume} {92}},\ \bibinfo
  {pages} {023002} (\bibinfo {year} {2015})},\ \Eprint
  {https://arxiv.org/abs/1502.04370} {arXiv:1502.04370 [gr-qc]} \BibitemShut
  {NoStop}%
\bibitem [{\citenamefont {Lange}\ \emph {et~al.}(2018)\citenamefont {Lange},
  \citenamefont {O'Shaughnessy},\ and\ \citenamefont {Rizzo}}]{Lange:2018pyp}%
  \BibitemOpen
  \bibfield  {author} {\bibinfo {author} {\bibfnamefont {J.}~\bibnamefont
  {Lange}}, \bibinfo {author} {\bibfnamefont {R.}~\bibnamefont
  {O'Shaughnessy}},\ and\ \bibinfo {author} {\bibfnamefont {M.}~\bibnamefont
  {Rizzo}},\ }\bibfield  {title} {\bibinfo {title} {{Rapid and accurate
  parameter inference for coalescing, precessing compact binaries}},\
  }\href@noop {} {\  (\bibinfo {year} {2018})},\ \Eprint
  {https://arxiv.org/abs/1805.10457} {arXiv:1805.10457 [gr-qc]} \BibitemShut
  {NoStop}%
\bibitem [{\citenamefont {Vinciguerra}\ \emph {et~al.}(2017)\citenamefont
  {Vinciguerra}, \citenamefont {Veitch},\ and\ \citenamefont
  {Mandel}}]{Vinciguerra:2017ngf}%
  \BibitemOpen
  \bibfield  {author} {\bibinfo {author} {\bibfnamefont {S.}~\bibnamefont
  {Vinciguerra}}, \bibinfo {author} {\bibfnamefont {J.}~\bibnamefont
  {Veitch}},\ and\ \bibinfo {author} {\bibfnamefont {I.}~\bibnamefont
  {Mandel}},\ }\bibfield  {title} {\bibinfo {title} {{Accelerating
  gravitational wave parameter estimation with multi-band template
  interpolation}},\ }\href {https://doi.org/10.1088/1361-6382/aa6d44}
  {\bibfield  {journal} {\bibinfo  {journal} {Class. Quant. Grav.}\ }\textbf
  {\bibinfo {volume} {34}},\ \bibinfo {pages} {115006} (\bibinfo {year}
  {2017})},\ \Eprint {https://arxiv.org/abs/1703.02062} {arXiv:1703.02062
  [gr-qc]} \BibitemShut {NoStop}%
\bibitem [{\citenamefont {Cannon}\ \emph {et~al.}(2012)\citenamefont {Cannon}
  \emph {et~al.}}]{Cannon:2011vi}%
  \BibitemOpen
  \bibfield  {author} {\bibinfo {author} {\bibfnamefont {K.}~\bibnamefont
  {Cannon}} \emph {et~al.},\ }\bibfield  {title} {\bibinfo {title} {{Toward
  Early-Warning Detection of Gravitational Waves from Compact Binary
  Coalescence}},\ }\href {https://doi.org/10.1088/0004-637X/748/2/136}
  {\bibfield  {journal} {\bibinfo  {journal} {Astrophys. J.}\ }\textbf
  {\bibinfo {volume} {748}},\ \bibinfo {pages} {136} (\bibinfo {year}
  {2012})},\ \Eprint {https://arxiv.org/abs/1107.2665} {arXiv:1107.2665
  [astro-ph.IM]} \BibitemShut {NoStop}%
\bibitem [{\citenamefont {Hannam}\ \emph {et~al.}(2014)\citenamefont {Hannam},
  \citenamefont {Schmidt}, \citenamefont {Boh\'e}, \citenamefont {Haegel},
  \citenamefont {Husa}, \citenamefont {Ohme}, \citenamefont {Pratten},\ and\
  \citenamefont {P\"urrer}}]{Hannam:2013oca}%
  \BibitemOpen
  \bibfield  {author} {\bibinfo {author} {\bibfnamefont {M.}~\bibnamefont
  {Hannam}}, \bibinfo {author} {\bibfnamefont {P.}~\bibnamefont {Schmidt}},
  \bibinfo {author} {\bibfnamefont {A.}~\bibnamefont {Boh\'e}}, \bibinfo
  {author} {\bibfnamefont {L.}~\bibnamefont {Haegel}}, \bibinfo {author}
  {\bibfnamefont {S.}~\bibnamefont {Husa}}, \bibinfo {author} {\bibfnamefont
  {F.}~\bibnamefont {Ohme}}, \bibinfo {author} {\bibfnamefont {G.}~\bibnamefont
  {Pratten}},\ and\ \bibinfo {author} {\bibfnamefont {M.}~\bibnamefont
  {P\"urrer}},\ }\bibfield  {title} {\bibinfo {title} {Simple model of complete
  precessing black-hole-binary gravitational waveforms},\ }\href
  {https://doi.org/10.1103/PhysRevLett.113.151101} {\bibfield  {journal}
  {\bibinfo  {journal} {Phys. Rev. Lett.}\ }\textbf {\bibinfo {volume} {113}},\
  \bibinfo {pages} {151101} (\bibinfo {year} {2014})}\BibitemShut {NoStop}%
\bibitem [{\citenamefont {Khan}\ \emph {et~al.}(2016)\citenamefont {Khan},
  \citenamefont {Husa}, \citenamefont {Hannam}, \citenamefont {Ohme},
  \citenamefont {P\"urrer}, \citenamefont {Forteza},\ and\ \citenamefont
  {Boh\'e}}]{Khan:2015jqa}%
  \BibitemOpen
  \bibfield  {author} {\bibinfo {author} {\bibfnamefont {S.}~\bibnamefont
  {Khan}}, \bibinfo {author} {\bibfnamefont {S.}~\bibnamefont {Husa}}, \bibinfo
  {author} {\bibfnamefont {M.}~\bibnamefont {Hannam}}, \bibinfo {author}
  {\bibfnamefont {F.}~\bibnamefont {Ohme}}, \bibinfo {author} {\bibfnamefont
  {M.}~\bibnamefont {P\"urrer}}, \bibinfo {author} {\bibfnamefont {X.~J.}\
  \bibnamefont {Forteza}},\ and\ \bibinfo {author} {\bibfnamefont
  {A.}~\bibnamefont {Boh\'e}},\ }\bibfield  {title} {\bibinfo {title}
  {{Frequency-domain gravitational waves from nonprecessing black-hole
  binaries. II. A phenomenological model for the advanced detector era}},\
  }\href {https://doi.org/10.1103/PhysRevD.93.044007} {\bibfield  {journal}
  {\bibinfo  {journal} {Phys. Rev.}\ }\textbf {\bibinfo {volume} {D93}},\
  \bibinfo {pages} {044007} (\bibinfo {year} {2016})},\ \Eprint
  {https://arxiv.org/abs/1508.07253} {arXiv:1508.07253 [gr-qc]} \BibitemShut
  {NoStop}%
\bibitem [{\citenamefont {Boh\'e}\ \emph {et~al.}(2016)\citenamefont {Boh\'e},
  \citenamefont {Hannam}, \citenamefont {Husa}, \citenamefont {Ohme},
  \citenamefont {P\"urrer},\ and\ \citenamefont {Schmidt}}]{Bohe:2016}%
  \BibitemOpen
  \bibfield  {author} {\bibinfo {author} {\bibfnamefont {A.}~\bibnamefont
  {Boh\'e}}, \bibinfo {author} {\bibfnamefont {M.}~\bibnamefont {Hannam}},
  \bibinfo {author} {\bibfnamefont {S.}~\bibnamefont {Husa}}, \bibinfo {author}
  {\bibfnamefont {F.}~\bibnamefont {Ohme}}, \bibinfo {author} {\bibfnamefont
  {M.}~\bibnamefont {P\"urrer}},\ and\ \bibinfo {author} {\bibfnamefont
  {P.}~\bibnamefont {Schmidt}},\ }\bibfield  {title} {\bibinfo {title}
  {{PhenomPv2 -- technical notes for the LAL implementation}},\ }\href
  {https://dcc.ligo.org/LIGO-T1500602/public} {\bibfield  {journal} {\bibinfo
  {journal} {LIGO Technical Document, LIGO-T1500602-v4}\ } (\bibinfo {year}
  {2016})}\BibitemShut {NoStop}%
\bibitem [{\citenamefont {Wildberger}\ \emph {et~al.}(2022)\citenamefont
  {Wildberger}, \citenamefont {Dax}, \citenamefont {Green}, \citenamefont
  {Gair}, \citenamefont {P\"urrer}, \citenamefont {Macke}, \citenamefont
  {Buonanno},\ and\ \citenamefont {Sch\"olkopf}}]{wildberger2022probabilistic}%
  \BibitemOpen
  \bibfield  {author} {\bibinfo {author} {\bibfnamefont {J.}~\bibnamefont
  {Wildberger}}, \bibinfo {author} {\bibfnamefont {M.}~\bibnamefont {Dax}},
  \bibinfo {author} {\bibfnamefont {S.~R.}\ \bibnamefont {Green}}, \bibinfo
  {author} {\bibfnamefont {J.}~\bibnamefont {Gair}}, \bibinfo {author}
  {\bibfnamefont {M.}~\bibnamefont {P\"urrer}}, \bibinfo {author}
  {\bibfnamefont {J.~H.}\ \bibnamefont {Macke}}, \bibinfo {author}
  {\bibfnamefont {A.}~\bibnamefont {Buonanno}},\ and\ \bibinfo {author}
  {\bibfnamefont {B.}~\bibnamefont {Sch\"olkopf}},\ }\bibfield  {title}
  {\bibinfo {title} {{Adapting to noise distribution shifts in flow-based
  gravitational-wave inference}},\ }\href@noop {} {\  (\bibinfo {year}
  {2022})},\ \Eprint {https://arxiv.org/abs/2211.08801} {arXiv:2211.08801
  [gr-qc]} \BibitemShut {NoStop}%
\bibitem [{\citenamefont {Payne}\ \emph {et~al.}(2022)\citenamefont {Payne},
  \citenamefont {Hourihane}, \citenamefont {Golomb}, \citenamefont {Udall},
  \citenamefont {Davis},\ and\ \citenamefont {Chatziioannou}}]{Payne:2022spz}%
  \BibitemOpen
  \bibfield  {author} {\bibinfo {author} {\bibfnamefont {E.}~\bibnamefont
  {Payne}}, \bibinfo {author} {\bibfnamefont {S.}~\bibnamefont {Hourihane}},
  \bibinfo {author} {\bibfnamefont {J.}~\bibnamefont {Golomb}}, \bibinfo
  {author} {\bibfnamefont {R.}~\bibnamefont {Udall}}, \bibinfo {author}
  {\bibfnamefont {D.}~\bibnamefont {Davis}},\ and\ \bibinfo {author}
  {\bibfnamefont {K.}~\bibnamefont {Chatziioannou}},\ }\bibfield  {title}
  {\bibinfo {title} {{Curious case of GW200129: Interplay between
  spin-precession inference and data-quality issues}},\ }\href
  {https://doi.org/10.1103/PhysRevD.106.104017} {\bibfield  {journal} {\bibinfo
   {journal} {Phys. Rev. D}\ }\textbf {\bibinfo {volume} {106}},\ \bibinfo
  {pages} {104017} (\bibinfo {year} {2022})},\ \Eprint
  {https://arxiv.org/abs/2206.11932} {arXiv:2206.11932 [gr-qc]} \BibitemShut
  {NoStop}%
\bibitem [{\citenamefont {Hannam}\ \emph {et~al.}(2022)\citenamefont {Hannam}
  \emph {et~al.}}]{Hannam:2021pit}%
  \BibitemOpen
  \bibfield  {author} {\bibinfo {author} {\bibfnamefont {M.}~\bibnamefont
  {Hannam}} \emph {et~al.},\ }\bibfield  {title} {\bibinfo {title}
  {{General-relativistic precession in a black-hole binary}},\ }\href
  {https://doi.org/10.1038/s41586-022-05212-z} {\bibfield  {journal} {\bibinfo
  {journal} {Nature}\ }\textbf {\bibinfo {volume} {610}},\ \bibinfo {pages}
  {652} (\bibinfo {year} {2022})},\ \Eprint {https://arxiv.org/abs/2112.11300}
  {arXiv:2112.11300 [gr-qc]} \BibitemShut {NoStop}%
\bibitem [{\citenamefont {Morras}\ \emph {et~al.}(2022)\citenamefont {Morras},
  \citenamefont {Siles}, \citenamefont {Garcia-Bellido},\ and\ \citenamefont
  {Morales}}]{Morras:2022ysx}%
  \BibitemOpen
  \bibfield  {author} {\bibinfo {author} {\bibfnamefont {G.}~\bibnamefont
  {Morras}}, \bibinfo {author} {\bibfnamefont {J.~F. N.~n.}\ \bibnamefont
  {Siles}}, \bibinfo {author} {\bibfnamefont {J.}~\bibnamefont
  {Garcia-Bellido}},\ and\ \bibinfo {author} {\bibfnamefont {E.~R.}\
  \bibnamefont {Morales}},\ }\bibfield  {title} {\bibinfo {title} {{The False
  Alarms induced by Gaussian Noise in Gravitational Wave Detectors}},\
  }\href@noop {} {\  (\bibinfo {year} {2022})},\ \Eprint
  {https://arxiv.org/abs/2209.05475} {arXiv:2209.05475 [gr-qc]} \BibitemShut
  {NoStop}%
\bibitem [{\citenamefont {Coughlin}\ \emph {et~al.}(2021)\citenamefont
  {Coughlin}, \citenamefont {Zevin}, \citenamefont {Bahaadini}, \citenamefont
  {Rohani}, \citenamefont {Allen}, \citenamefont {Berry}, \citenamefont
  {Crowston}, \citenamefont {Harandi}, \citenamefont {Jackson}, \citenamefont
  {Kalogera}, \citenamefont {Katsaggelos}, \citenamefont {Noroozi},
  \citenamefont {Osterlund}, \citenamefont {Patane}, \citenamefont {Smith},
  \citenamefont {Soni},\ and\ \citenamefont
  {Trouille}}]{coughlin_scott_2021_5649212}%
  \BibitemOpen
  \bibfield  {author} {\bibinfo {author} {\bibfnamefont {S.}~\bibnamefont
  {Coughlin}}, \bibinfo {author} {\bibfnamefont {M.}~\bibnamefont {Zevin}},
  \bibinfo {author} {\bibfnamefont {S.}~\bibnamefont {Bahaadini}}, \bibinfo
  {author} {\bibfnamefont {N.}~\bibnamefont {Rohani}}, \bibinfo {author}
  {\bibfnamefont {S.}~\bibnamefont {Allen}}, \bibinfo {author} {\bibfnamefont
  {C.}~\bibnamefont {Berry}}, \bibinfo {author} {\bibfnamefont
  {K.}~\bibnamefont {Crowston}}, \bibinfo {author} {\bibfnamefont
  {M.}~\bibnamefont {Harandi}}, \bibinfo {author} {\bibfnamefont
  {C.}~\bibnamefont {Jackson}}, \bibinfo {author} {\bibfnamefont
  {V.}~\bibnamefont {Kalogera}}, \bibinfo {author} {\bibfnamefont
  {A.}~\bibnamefont {Katsaggelos}}, \bibinfo {author} {\bibfnamefont
  {V.}~\bibnamefont {Noroozi}}, \bibinfo {author} {\bibfnamefont
  {C.}~\bibnamefont {Osterlund}}, \bibinfo {author} {\bibfnamefont
  {O.}~\bibnamefont {Patane}}, \bibinfo {author} {\bibfnamefont
  {J.}~\bibnamefont {Smith}}, \bibinfo {author} {\bibfnamefont
  {S.}~\bibnamefont {Soni}},\ and\ \bibinfo {author} {\bibfnamefont
  {L.}~\bibnamefont {Trouille}},\ }\bibfield  {title} {\bibinfo {title}
  {{Gravity Spy Machine Learning Classifications of LIGO Glitches from
  Observing Runs O1, O2, O3a, and O3b}},\ }\href
  {https://doi.org/10.5281/zenodo.5649212} {10.5281/zenodo.5649212} (\bibinfo
  {year} {2021})\BibitemShut {NoStop}%
\bibitem [{\citenamefont {Szegedy}\ \emph {et~al.}(2014)\citenamefont
  {Szegedy}, \citenamefont {Zaremba}, \citenamefont {Sutskever}, \citenamefont
  {Bruna}, \citenamefont {Erhan}, \citenamefont {Goodfellow},\ and\
  \citenamefont {Fergus}}]{szegedy2013intriguing}%
  \BibitemOpen
  \bibfield  {author} {\bibinfo {author} {\bibfnamefont {C.}~\bibnamefont
  {Szegedy}}, \bibinfo {author} {\bibfnamefont {W.}~\bibnamefont {Zaremba}},
  \bibinfo {author} {\bibfnamefont {I.}~\bibnamefont {Sutskever}}, \bibinfo
  {author} {\bibfnamefont {J.}~\bibnamefont {Bruna}}, \bibinfo {author}
  {\bibfnamefont {D.}~\bibnamefont {Erhan}}, \bibinfo {author} {\bibfnamefont
  {I.}~\bibnamefont {Goodfellow}},\ and\ \bibinfo {author} {\bibfnamefont
  {R.}~\bibnamefont {Fergus}},\ }\bibfield  {title} {\bibinfo {title}
  {Intriguing properties of neural networks},\ }in\ \href@noop {} {\emph
  {\bibinfo {booktitle} {International Conference on Learning
  Representations}}}\ (\bibinfo {year} {2014})\ \Eprint
  {https://arxiv.org/abs/1312.6199} {arXiv:1312.6199 [cs.LG]} \BibitemShut
  {NoStop}%
\bibitem [{\citenamefont {Goodfellow}\ \emph {et~al.}(2015)\citenamefont
  {Goodfellow}, \citenamefont {Shlens},\ and\ \citenamefont
  {Szegedy}}]{goodfellow2014explaining}%
  \BibitemOpen
  \bibfield  {author} {\bibinfo {author} {\bibfnamefont {I.~J.}\ \bibnamefont
  {Goodfellow}}, \bibinfo {author} {\bibfnamefont {J.}~\bibnamefont {Shlens}},\
  and\ \bibinfo {author} {\bibfnamefont {C.}~\bibnamefont {Szegedy}},\
  }\bibfield  {title} {\bibinfo {title} {Explaining and harnessing adversarial
  examples},\ }in\ \href@noop {} {\emph {\bibinfo {booktitle} {International
  Conference on Learning Representations}}}\ (\bibinfo {year} {2015})\ \Eprint
  {https://arxiv.org/abs/1412.6572} {arXiv:1412.6572 [cs.LG]} \BibitemShut
  {NoStop}%
\bibitem [{\citenamefont {Paszke}\ \emph {et~al.}(2019)\citenamefont {Paszke},
  \citenamefont {Gross}, \citenamefont {Massa}, \citenamefont {Lerer},
  \citenamefont {Bradbury}, \citenamefont {Chanan}, \citenamefont {Killeen},
  \citenamefont {Lin}, \citenamefont {Gimelshein}, \citenamefont {Antiga},
  \citenamefont {Desmaison}, \citenamefont {Kopf}, \citenamefont {Yang},
  \citenamefont {DeVito}, \citenamefont {Raison}, \citenamefont {Tejani},
  \citenamefont {Chilamkurthy}, \citenamefont {Steiner}, \citenamefont {Fang},
  \citenamefont {Bai},\ and\ \citenamefont {Chintala}}]{NEURIPS2019_9015}%
  \BibitemOpen
  \bibfield  {author} {\bibinfo {author} {\bibfnamefont {A.}~\bibnamefont
  {Paszke}}, \bibinfo {author} {\bibfnamefont {S.}~\bibnamefont {Gross}},
  \bibinfo {author} {\bibfnamefont {F.}~\bibnamefont {Massa}}, \bibinfo
  {author} {\bibfnamefont {A.}~\bibnamefont {Lerer}}, \bibinfo {author}
  {\bibfnamefont {J.}~\bibnamefont {Bradbury}}, \bibinfo {author}
  {\bibfnamefont {G.}~\bibnamefont {Chanan}}, \bibinfo {author} {\bibfnamefont
  {T.}~\bibnamefont {Killeen}}, \bibinfo {author} {\bibfnamefont
  {Z.}~\bibnamefont {Lin}}, \bibinfo {author} {\bibfnamefont {N.}~\bibnamefont
  {Gimelshein}}, \bibinfo {author} {\bibfnamefont {L.}~\bibnamefont {Antiga}},
  \bibinfo {author} {\bibfnamefont {A.}~\bibnamefont {Desmaison}}, \bibinfo
  {author} {\bibfnamefont {A.}~\bibnamefont {Kopf}}, \bibinfo {author}
  {\bibfnamefont {E.}~\bibnamefont {Yang}}, \bibinfo {author} {\bibfnamefont
  {Z.}~\bibnamefont {DeVito}}, \bibinfo {author} {\bibfnamefont
  {M.}~\bibnamefont {Raison}}, \bibinfo {author} {\bibfnamefont
  {A.}~\bibnamefont {Tejani}}, \bibinfo {author} {\bibfnamefont
  {S.}~\bibnamefont {Chilamkurthy}}, \bibinfo {author} {\bibfnamefont
  {B.}~\bibnamefont {Steiner}}, \bibinfo {author} {\bibfnamefont
  {L.}~\bibnamefont {Fang}}, \bibinfo {author} {\bibfnamefont {J.}~\bibnamefont
  {Bai}},\ and\ \bibinfo {author} {\bibfnamefont {S.}~\bibnamefont
  {Chintala}},\ }\bibfield  {title} {\bibinfo {title} {Pytorch: An imperative
  style, high-performance deep learning library},\ }in\ \href
  {http://papers.neurips.cc/paper/9015-pytorch-an-imperative-style-high-performance-deep-learning-library.pdf}
  {\emph {\bibinfo {booktitle} {Advances in Neural Information Processing
  Systems 32}}},\ \bibinfo {editor} {edited by\ \bibinfo {editor}
  {\bibfnamefont {H.}~\bibnamefont {Wallach}}, \bibinfo {editor} {\bibfnamefont
  {H.}~\bibnamefont {Larochelle}}, \bibinfo {editor} {\bibfnamefont
  {A.}~\bibnamefont {Beygelzimer}}, \bibinfo {editor} {\bibfnamefont
  {F.}~\bibnamefont {d'Alch\'{e} Buc}}, \bibinfo {editor} {\bibfnamefont
  {E.}~\bibnamefont {Fox}},\ and\ \bibinfo {editor} {\bibfnamefont
  {R.}~\bibnamefont {Garnett}}}\ (\bibinfo  {publisher} {Curran Associates,
  Inc.},\ \bibinfo {year} {2019})\ pp.\ \bibinfo {pages}
  {8024--8035}\BibitemShut {NoStop}%
\bibitem [{\citenamefont {Durkan}\ \emph {et~al.}(2020)\citenamefont {Durkan},
  \citenamefont {Bekasov}, \citenamefont {Murray},\ and\ \citenamefont
  {Papamakarios}}]{nflows}%
  \BibitemOpen
  \bibfield  {author} {\bibinfo {author} {\bibfnamefont {C.}~\bibnamefont
  {Durkan}}, \bibinfo {author} {\bibfnamefont {A.}~\bibnamefont {Bekasov}},
  \bibinfo {author} {\bibfnamefont {I.}~\bibnamefont {Murray}},\ and\ \bibinfo
  {author} {\bibfnamefont {G.}~\bibnamefont {Papamakarios}},\ }\href
  {https://doi.org/10.5281/zenodo.4296287} {\bibinfo {title} {{nflows}:
  normalizing flows in {PyTorch}}} (\bibinfo {year} {2020})\BibitemShut
  {NoStop}%
\bibitem [{\citenamefont {{LIGO Scientific Collaboration}}(2018)}]{lalsuite}%
  \BibitemOpen
  \bibfield  {author} {\bibinfo {author} {\bibnamefont {{LIGO Scientific
  Collaboration}}},\ }\href {https://doi.org/10.7935/GT1W-FZ16} {\bibinfo
  {title} {{LIGO} {A}lgorithm {L}ibrary - {LALS}uite}},\ \bibinfo
  {howpublished} {free software (GPL)} (\bibinfo {year} {2018})\BibitemShut
  {NoStop}%
\bibitem [{\citenamefont {Kingma}\ and\ \citenamefont
  {Ba}(2015)}]{Kingma:2014vow}%
  \BibitemOpen
  \bibfield  {author} {\bibinfo {author} {\bibfnamefont {D.~P.}\ \bibnamefont
  {Kingma}}\ and\ \bibinfo {author} {\bibfnamefont {J.}~\bibnamefont {Ba}},\
  }\bibfield  {title} {\bibinfo {title} {{Adam: A Method for Stochastic
  Optimization}},\ }in\ \href@noop {} {\emph {\bibinfo {booktitle}
  {International Conference on Learning Representations}}}\ (\bibinfo {year}
  {2015})\ \Eprint {https://arxiv.org/abs/1412.6980} {arXiv:1412.6980 [cs.LG]}
  \BibitemShut {NoStop}%
\bibitem [{\citenamefont {Hunter}(2007)}]{Hunter:2007}%
  \BibitemOpen
  \bibfield  {author} {\bibinfo {author} {\bibfnamefont {J.~D.}\ \bibnamefont
  {Hunter}},\ }\bibfield  {title} {\bibinfo {title} {Matplotlib: A 2d graphics
  environment},\ }\href {https://doi.org/10.1109/MCSE.2007.55} {\bibfield
  {journal} {\bibinfo  {journal} {Computing in Science \& Engineering}\
  }\textbf {\bibinfo {volume} {9}},\ \bibinfo {pages} {90} (\bibinfo {year}
  {2007})}\BibitemShut {NoStop}%
\bibitem [{\citenamefont {{Hinton}}(2016)}]{Hinton2016}%
  \BibitemOpen
  \bibfield  {author} {\bibinfo {author} {\bibfnamefont {S.~R.}\ \bibnamefont
  {{Hinton}}},\ }\bibfield  {title} {\bibinfo {title} {{ChainConsumer}},\
  }\href {https://doi.org/10.21105/joss.00045} {\bibfield  {journal} {\bibinfo
  {journal} {The Journal of Open Source Software}\ }\textbf {\bibinfo {volume}
  {1}},\ \bibinfo {eid} {00045} (\bibinfo {year} {2016})}\BibitemShut {NoStop}%
\end{thebibliography}%

\clearpage
\begin{center}
  \large
  \textbf{Supplemental Material}
\end{center}
\section{Importance-sampled Bayesian evidence}

The Bayesian evidence is given by
\begin{equation}\label{eq:evidence-is}
    p(d) = \int d\theta p(d|\theta)p(\theta) = \int d\theta \frac{p(d|\theta)p(\theta)}{q(\theta|d)} q(\theta|d),
\end{equation}
which can be estimated using $n$ samples $\theta_i\sim q(\theta|d)$ in the Monte Carlo approximation as $p(d) = \hat{\mu}_{w}$ with
\begin{equation}
    \hat\mu_w = \frac{1}{n} \sum_{i} \frac{p(d|\theta_i)p(\theta_i)}{q(\theta_i|d)} = 
    \frac{1}{n} \sum_{i} w_i
\end{equation}
where $w_i={p(d|\theta_i)p(\theta_i)}/{q(\theta_i|d)}$ are the weights used for importance sampling. The variance for this Monte Carlo estimate is given by 
\begin{equation}
\begin{split}
    \sigma_w^2 & = \text{Var}\left[\frac{p(d|\theta)p(\theta)}{q(\theta|d)}\right]
    \approx \frac{1}{n} \sum_i \left(w_i - \hat\mu_w \right)^2\\
    &= \hat\mu_w^2 \cdot \frac{1}{n} \sum_i \left[\bar w_i - 1 \right]^2
    = \hat\mu_w^2 \cdot \left(\frac{1}{n}\sum_i \bar w_i^2 -1\right)\\
    &
    = \hat\mu_w^2 \cdot \left(\frac{n - n_\text{eff}}{n_\text{eff}}\right)
    = \hat\mu_w^2 \cdot \left(\frac{1 - \epsilon}{\epsilon}\right),
\end{split}
\end{equation}
where we denote normalized weights with $\bar w_i=w_i/\hat\mu_w$ and the sample efficiency with $\epsilon = n_\text{eff}/n$. Since we use $n$ samples to estimate $p(d) = \hat{\mu}_{w}$, the standard deviation of the evidence is given by 
\begin{equation}
    \sigma_{p(d)} = \frac{\sigma_w}{\sqrt{n}} 
    = p(d) \sqrt{\frac{1 - \epsilon}{n\cdot\epsilon}}.
\end{equation}
In practice, we are interested in the log evidence, for which the uncertainty is
\begin{equation}
    \sigma_{\log p(d)} = \frac{\sigma_{p(d)}}{p(d)} 
    = \sqrt{\frac{1 - \epsilon}{n\cdot\epsilon}}.
  \end{equation}
\subsection{Bias}
Since $p(\theta)$ and $q(\theta|d)$ are normalized, Eq.~\eqref{eq:evidence-is} provides an unbiased estimate for $p(d)$~\cite{mcbook},
\begin{equation}
    \mathbb{E}\left[\frac{1}{n} \sum_{i} w_i\right] = \mathbb{E}[\hat\mu_w] = p(d).
\end{equation}
The logarithm of the evidence however has a bias. Defining $Y = \hat\mu_w - p(d)$, we find
\begin{equation}
    \begin{split}
    \mathbb{E}[\log\hat\mu_w] 
    & = \mathbb{E}\left[\log\left(p(d) + p(d)\cdot \frac{\hat\mu_w - p(d)}{p(d)}\right)\right]\\
    & = \log p(d) + \mathbb{E}\left[\log\left(1 + \frac{Y}{p(d)}\right)\right]\\
    & = \log p(d) + \mathbb{E}\left[\frac{Y}{p(d)} - \frac{1}{2}\left(\frac{Y}{p(d)}\right)^2 \right]\\
    & = \log p(d) - \frac{\sigma_w^2}{2 p(d)^2 n}
    = \log p(d) - \frac{1-\epsilon}{2n\epsilon}
    \end{split}
\end{equation}
where we used $\mathbb{E}[Y] = 0$ and $\text{Var}[Y] = \sigma_w^2/n$ and neglected terms of order $\mathcal{O}((Y/p(d))^3)$. The bias of the log evidence thus depends on the sample efficiency $\epsilon = n_\text{eff}/n$ and scales with $1/n$. Given that the uncertainty of $\log\hat\mu_w$ scales with $1/\sqrt{n}$, this bias is completely negligible in practice.

\section{Analytic estimate of the phase parameter}

The parameter $\phi_\text{c}$ describes the phase of the gravitational wave at a fixed reference frequency. It provides no physical insight, but it is necessary to define a complete likelihood~\cite{veitch2013analytic}. While the marginal $p(\phi_\text{c}|d)$ usually has a simple structure, the \emph{conditional} distribution $p(\phi_\text{c}|d,\tilde\theta)$, where $\tilde\theta$ denotes the 14 remaining parameters, is typically very tightly constrained. Furthermore, $\phi_\text{c}$ is strongly correlated with $\tilde\theta$. We observed that \textsc{Dingo} has difficulties learning the phase parameter, and often infers the prior instead, $q(\phi_\text{c}|d,\tilde\theta) = p(\phi_\text{c})$. While we did not find this to have a negative impact on the remaining parameters, it leads to a substantially reduced sample efficiency.

Inspired by phase marginalization~\cite{veitch2013analytic,Thrane:2018qnx}, a technique commonly used to increase the efficiency of stochastic samplers, we analytically estimate $\phi_\text{c}$. The approach outlined below differs in two ways from typical phase marginalization---(1) we retrieve $\phi_\text{c}$ instead of marginalizing over it, and (2) this technique is exact even in the presence of higher modes, where phase marginalization is an approximation. 

We decompose our posterior estimate into
\begin{equation}
    q(\theta|d) = p(\phi_\text{c}|d, \tilde\theta) q(\tilde\theta|d),
\end{equation}
where $q(\tilde\theta|d)$ is estimated with \textsc{Dingo}. For each \textsc{Dingo} sample $\tilde\theta\sim q(\tilde\theta|d)$, we then \emph{synthetically} sample $\phi_\text{c}$ using the analytic likelihood. This is done by evaluating $p(\phi_\text{c}|d, \tilde\theta)$ on a uniform grid over $\phi_\text{c}$ with 5001 points in the range $[0,2\pi]$ and interpolating in between. 

Each likelihood evaluation requires a waveform simulation, which accounts for the bulk of the computational cost. As we outline below, by caching suitable combinations of the waveform modes, we can cheaply evaluate waveform polarizations for arbitrary $\phi_c$. Hence sampling the synthetic $\phi_\text{c}$ is barely more expensive than a single likelihood evaluation.

\subsection{Phase transformations}

We work in the $L_0$ frame, which aligns the $z$ axis with the orbital angular momentum of the binary at the reference frequency, and takes $\phi_c$ as the azimuthal angle of the observer relative to the axis connecting the two bodies. In these coordinates, the observer is located at $(\theta, \phi) = (\iota, \pi/2 - \phi_c)$, where $\iota$ is the inclination of the binary. This is convenient for caching the modes, since $\phi_c$ enters the waveform entirely via the spin-weighted spherical harmonics (as opposed to the modes themselves).

Waveform modes $h_{\ell m}$ combine into polarizations $h_{+,\times}$ as
\begin{equation}
    h_+ - i h_\times = h = \sum_{\ell,m} h_{\ell m} \, {}_{-2}Y_{\ell m}(\theta, \phi),
\end{equation}
In frequency domain,
\begin{eqnarray}
    \tilde h_+(f) &=& \frac{1}{2}\left[ \tilde h(f) + \tilde h^\ast(-f) \right], \\
    \tilde h_\times(f) &=& \frac{i}{2}\left[ \tilde h(f) - \tilde h^\ast(-f) \right].
\end{eqnarray}
Considering just the plus polarization and substituting for the mode expansion,
\begin{eqnarray}
    \tilde h_+(f) &=& \frac{1}{2}\sum_{\ell, m} \left[ \tilde h_{\ell m} (f) \, {}_{-2}Y_{\ell m}(\theta, \phi) \right. \nonumber \\
    && \qquad \quad + \left.\tilde h_{\ell m}^\ast(-f)\, {}_{-2}Y^\ast_{\ell m}(\theta, \phi) \right].
\end{eqnarray}
Now we use the fact that the $\phi$-dependence enters the spin-weighted spherical harmonics as ${}_{-2}Y_{\ell m}(\theta, \phi) = {}_{-2}Y_{\ell m}(\theta, 0) e^{i m \phi}$. Since $h_+$ is real, we only need to consider $f>0$. In the $L_0$ frame, we can then write
\begin{equation}
    \tilde h_+(f > 0) = \sum_{m} \tilde h_{+,m}(f) e^{- i m \phi_c}, 
\end{equation}
where we have grouped the terms according to their $m$-dependence,
\begin{eqnarray}
    \tilde h_{+,m}(f) &=& \frac{1}{2} \sum_\ell \left[ \tilde h_{\ell m}(f) \, {}_{-2}Y_{\ell m}\left(\iota, \frac{\pi}{2}\right) \right. \nonumber\\ 
    && \qquad + \left. \tilde h_{\ell, -m}^\ast(-f) \, {}_{-2}Y_{\ell, -m}^\ast\left(\iota, \frac{\pi}{2}\right)\right].
\end{eqnarray}
Notice that we combined the positive frequency parts of modes with azimuthal number $m$ together with negative frequency modes of azimuthal number $-m$. With this decomposition, we only need to cache the $\tilde h_{+,m}$. Likewise for the cross polarization, we have
\begin{equation}
    \tilde h_\times(f > 0) = \sum_{m} \tilde h_{\times,m}(f) e^{- i m \phi_c}, 
\end{equation}
where
\begin{eqnarray}
    h_{\times,m}(f) &=& \frac{i}{2} \sum_\ell \left[ \tilde h_{\ell m}(f) \, {}_{-2}Y_{\ell m}\left(\iota, \frac{\pi}{2}\right) \right. \nonumber\\ 
    && \qquad - \left. \tilde h_{\ell, -m}^\ast(-f) \, {}_{-2}Y_{\ell, -m}^\ast\left(\iota, \frac{\pi}{2}\right)\right].
\end{eqnarray}

One additional complication arises because waveform models are usually given in terms of Cartesian spin components, and $\phi_c$ also enters into their definition in terms of the spin parameters used for parameter estimation. Consequently the modes retain a dependence on $\phi_c$. We overcome this by fixing the phase parameter used in effecting this transformation. This results in a slightly different definition of the spin parameters $\theta_{JN}$ and $\phi_{JL}$, which we undo in post-processing. Since the standard priors are invariant under this transformation, other parameters are not affected.

This approach enables likelihood evaluations on a $\phi_\text{c}$ grid at the computational cost of a single likelihood evaluation, plus a small additional cost for the inner products.\footnote{For IMRPhenomXPHM, computing the individual modes with the \textsc{LALSimulation} function \texttt{SimInspiralChooseFDModes} is substantially more expensive than computing the combined polarizations with \texttt{SimInspiralFD}. This is because \texttt{SimInspiralFD} caches information when internally computing the modes, whereas \texttt{SimInspiralChooseFDModes} does not.} The implementation is fully contained in the \textsc{Dingo} package, which uses low-level LALSimulation~\cite{lalsuite} functions to compute frequency domain modes in the $L_0$ frame, and combines them into the $\tilde h_{+/\times,m}$. For SEOBNRv4PHM, this requires Fourier transforming the time domain modes provided by \textsc{LALSimulation} in $L_0$ frame. For IMRPhenomXPHM it requires transforming from $J$ to $L_0$ frame, such that the $\phi_\text{c}$ dependence enters via the spherical harmonics, not via the modes themselves.

\section{Density recovery}
IS requires access to the density of the inferred samples. While for NPE, this density is tractable, this is not necessarily the case for other inference methods. Below, we describe how we use neural density estimation to recover the density in these cases.

\subsection{Group equivariant neural posterior estimation}
\textsc{Dingo} uses an iterative algorithm called \emph{group equivariant} NPE (GNPE)~\mbox{\cite{Dax:2021tsq,Dax:2021myb}} to integrate physical symmetries and thereby improve the accuracy of inference. With GNPE, we train a density estimation network $q(\theta|d,\hat{t}_I(\theta))$ that is also conditional on a set of GNPE \emph{proxy parameters} $\hat{t}_I$. These parameters are defined as blurred versions of the coalescence times $t_I$ in the individual interferometers (which can be computed as a function of $\theta$) as
\begin{equation}
    \hat t_I = t_I + \epsilon_I,\quad \epsilon_I\sim \kappa(\epsilon),
\end{equation}
with $\kappa = U[-1~\text{ms}, 1\text{ms}]$. With GNPE, we iteratively infer the posterior $p(\theta,\hat t_I|d)$ in the joint parameter space with Gibbs sampling, and obtain the posterior over $\theta$ by marginalizing over $\hat t_I$. We use a paralllelized Gibbs sampler that typically converges after 30 iterations, but some events require up to 500 iterations. Each iteration corresponds to a forward pass through the density estimator $q(\theta|d,\hat{t}_I(\theta))$. 500 GNPE iterations for a batch of $5\cdot 10^4$ samples take about 6~minutes on an A100 GPU.

\begin{table}[]
    \centering
    \begin{tabular}{lrr}
        \hline\hline
        & ~~2 dimensions & ~~14 dimensions\\\hline
        flow steps & 5 & 20\\
        hidden dimension & 256 & 256\\
        transform blocks & 4 & 4\\
        bins & 8 & 8\\\hline
        training samples & $4\cdot 10^5$ & $10^6$\\
        batch size & 4096 & 8192\\
        epochs & 20 & 60\\
        optimizer & adam~\cite{Kingma:2014vow} & adam~\cite{Kingma:2014vow}\\
        learning rate & 0.002 & 0.001\\
        training time on A100 GPU & 7~minutes & 1~hour\\\hline\hline
    \end{tabular}
    \caption{Settings for the neural spline flow~\cite{durkan2019neural} architecture (upper part) and training (lower) used for density recovery. For \textsc{Dingo-IS} with GNPE, we need to estimate a two dimensional distribution over the proxy parameters, which requires a smaller network than the distribution over the 14 dimensional parameter space used for \textsc{Bilby-IS}.}
    \label{tab:unconditional-flows-hyperparams}
\end{table}

In contrast to NPE, GNPE does not have a tractable density. To recover the density, we first generate $4\cdot 10^5$ GNPE samples (48~minutes on one GPU or 6~minutes on eight GPUs for 500 iterations). We then train an unconditional normalizing flow $q(\hat t_I)$ to estimate the distribution over the inferred proxy parameters with a maximum likelihood objective. We use a neural spline flow with rational-quadratic spline coupling transforms~\cite{durkan2019neural} with the hyperparameters from Tab.~\ref{tab:unconditional-flows-hyperparams}. Once trained, we can sample without the need for additional GNPE iterations via
\begin{equation}
    \theta\sim q(\theta|d, \hat t_I),~\hat t_I\sim q(\hat t_I).
\end{equation}
The proposal density is now tractable,
\begin{equation}
    \log q(\theta,\hat t_I|d) = \log q(\theta|d, \hat t_I) + \log q(\hat t_I).
\end{equation}
We then perform IS in the \emph{joint} parameter space $(\theta, \hat t_I)$, where the target density is given by
\begin{equation}
    \begin{split}
    \log p(\theta,\hat t_I|d) = 
    -\log p(d) &+ \log p(d|\theta) + \log p(\theta) \\&+ \sum_I \log \kappa(\hat t_I - t_I).
    \end{split}
\end{equation}
The last term accounts for $p(\hat t_I|\theta)$. As described in the main part, we omit $\log p(d)$ and estimate this from the normalization of the weights. 

Alternatively, we could also train an unconditional density estimator for the converged $\theta$ samples, but this is less sample efficient and more costly to train.

\subsection{Stochastic samplers}

We apply IS to \textsc{Bilby-dynesty}~\mbox{\cite{Ashton:2018jfp,Romero-Shaw:2020owr,Speagle_2020}}, which is based on nested sampling. To recover the density, we first generate $\approx 10^6$ posterior samples with 50 \textsc{Bilby} runs with identical settings. With \texttt{nlive=1000} and \texttt{nact=5}, this takes about one day per run on 10 CPUs, when using the IMRPhenomXPHM model. One typically uses larger \texttt{nact} for production results, but this substantially increases the computational cost. For reference, the runs for GW150914 and GW151012 reported in the main paper with \texttt{nlive=4000} and \texttt{nact=50} took about a week. We then estimate the distribution over the \textsc{Bilby} samples by training an unconditional normalizing flow $q(\theta)$, see Tab.~\ref{tab:unconditional-flows-hyperparams}. To ensure a fair comparison with \textsc{Dingo}, we also use the analytic estimate for the phase parameter, such that we only need to estimate the distribution over the remaining 14 parameters. Due to the higher dimensional parameter space compared to \textsc{Dingo} (for which we only need to recover the two dimensional density over $\hat t_I$), we need more samples and a larger normalizing flow for the density estimate.

For GW151012, \textsc{Bilby}-IS achieves a sample efficiency $\epsilon = 8.3\%$, compared to $\epsilon = 12.5\%$ for \textsc{Dingo}-IS, and estimates an evidence of $\log p(d) = -16412.89\pm 0.01$. Since \textsc{Bilby-IS} is computationally very expensive, we do not expect it to be routinely used, but rather view it as an insightful diagnostic.

\section{Importance sampling convergence}

\begin{figure}
  \includegraphics[width=0.48\textwidth]{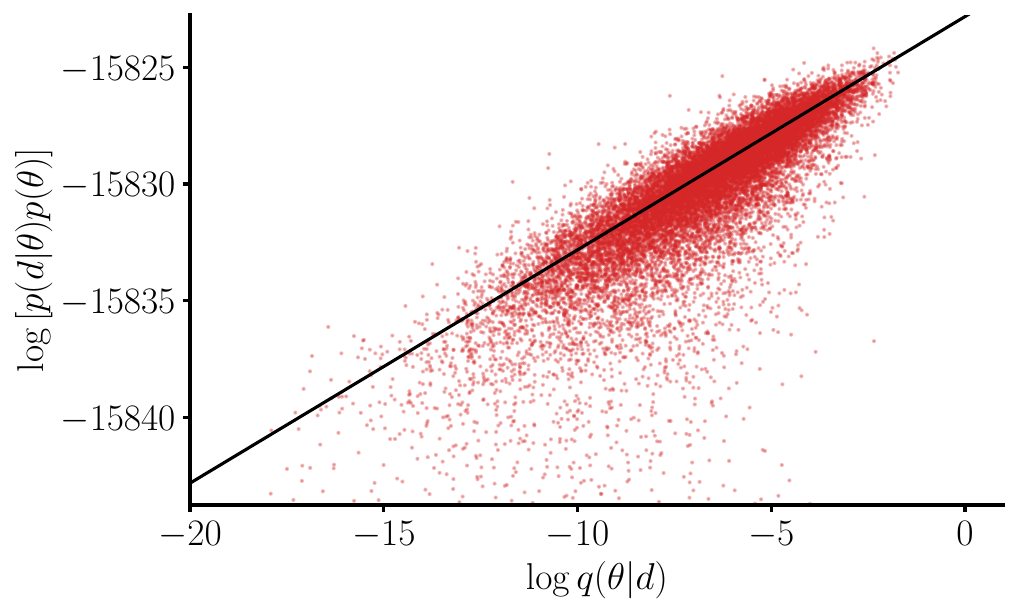}
  \caption{\label{fig:log-densities-GW150914}\textsc{Dingo} samples $\theta\sim q(\theta|d)$ for GW150914, comparing the inferred density $q(\theta|d)$ to the unnormalized posterior $p(d|\theta)p(\theta)$. The density ratios correspond to the importance weights, the Bayesian evidence $p(d)$ is estimated via their normalization. Samples of a perfect \textsc{Dingo} model would lie on the black line with offset $\log p(d)=-15831.87$. Deviations between \textsc{Dingo} and the true posterior are primarily found below that line, but rarely above. This is a manifestation of the probability-mass covering behavior, making \textsc{Dingo} particularly well-suited for importance sampling.
  }
\end{figure}

\begin{figure*}
  \includegraphics[width=0.49\textwidth]{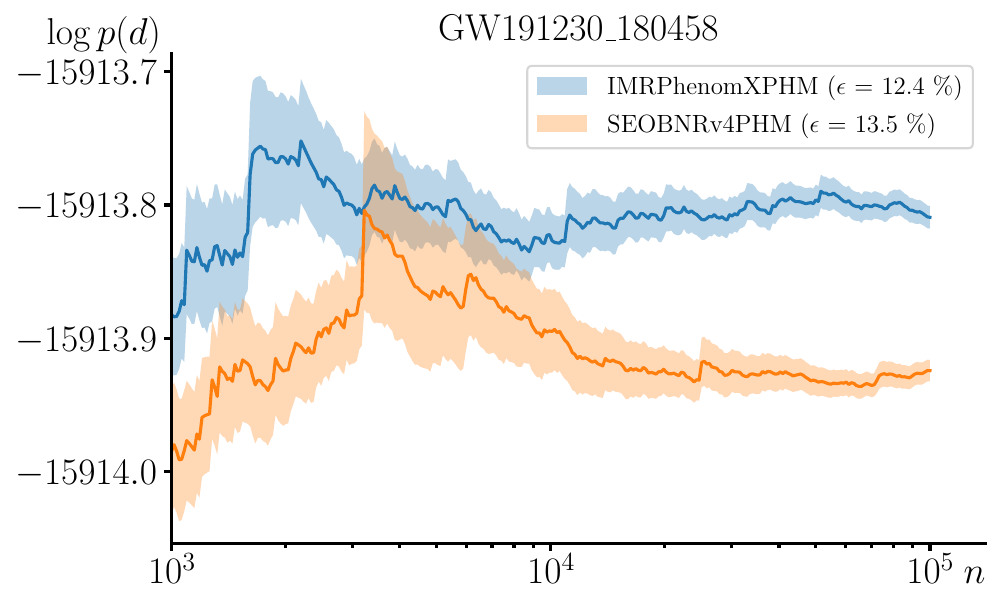}
  \hfill
  \includegraphics[width=0.49\textwidth]{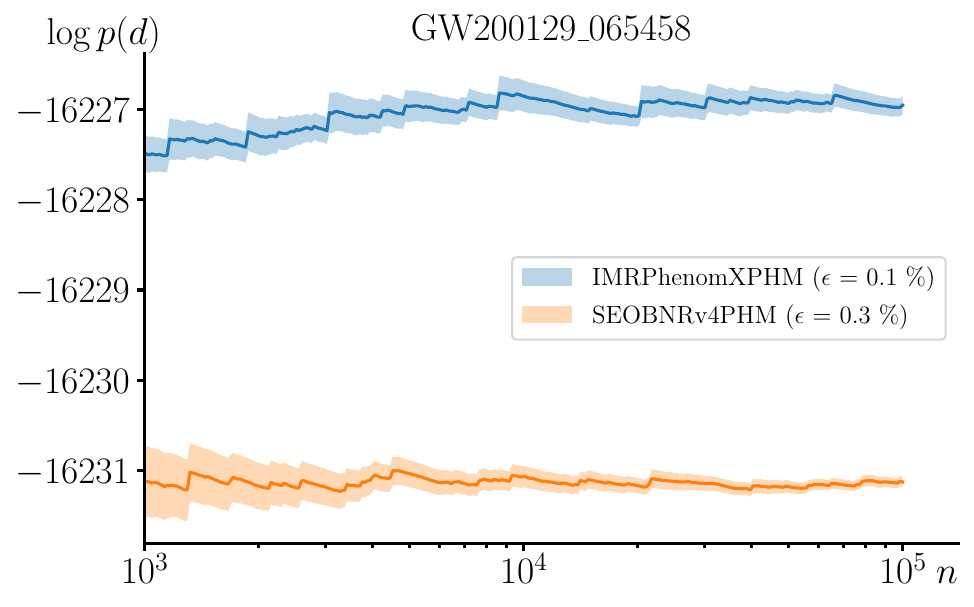}
  \caption{\label{fig:evidence-vs-n}
  Evidence $\log p(d)$ as a function of the number of importance samples $n$. For constant sample efficiency $\epsilon$, the statistical uncertainty scales with $1/\sqrt{n}$, leading to precise estimates when \textsc{Dingo-IS} works well (left). When the \textsc{Dingo} posterior is too light tailed (right), samples from the tails of the distribution are assigned very large IS weights, leading to bumps in the evidence whenever a high-weight sample is encountered.
  }
\end{figure*}

Due to the probability mass covering training objective, \textsc{Dingo} inaccuracies tend to show up as overly broad posteriors (Fig.~\ref{fig:log-densities-GW150914}). When the tails of the posterior are overestimated by \textsc{Dingo}, a low sample efficiency may be encountered due to many low-weight samples. These cases are straightforward to handle with \textsc{Dingo-IS}. The sample efficiency is approximately constant, and the statistical uncertainty of the evidence fully captures the error, even for low $n_\text{eff}$. To get smooth marginals one simply needs to generate more samples, which is cheap with \textsc{Dingo}.

In contrast, \textsc{Dingo} posteriors should rarely be light tailed. For real data, however, parts of the parameter space are occasionally strongly undersampled, which is problematic for IS. Indeed, for small $n$, the light tails may not be sampled at all, resulting in an underestimate of the evidence and the magnitude of its statistical error. Moreover, when a sample from the tail is encountered it has very large importance weight, which greatly decreases the sample efficiency. In order to assess the validity of IS results with low sample efficiency it is therefore useful to check whether $\log p(d)$ has converged as a function of $n$ (Fig.~\ref{fig:evidence-vs-n}). If \textsc{Dingo} is not truly mass covering, the IS weights are not upper-bounded, and the sample efficiency approaches zero with increasing $n$. This happens for the OOD event GW200129\_065458.

Fortunately non-convergence is rare, and for the majority of events, \textsc{Dingo} posteriors are indeed mass covering and heavy tailed. Even when this is not the case and the sample efficiency is very low, the \textsc{Dingo} marginals are often still accurate. This is because the light tailed parts of the parameter space are often negligibly small and randomly distributed throughout the parameter space. In such cases one can apply \emph{batched} self-normalized IS: instead of normalizing the weights of all $n$ samples simultaneously, one normalizes batches of size $k<n$. This regularizes IS by decreasing the largest possible weight from $n$ to $k$. This should be done with caution, as it introduces a bias which is only small if the undersampled regions carry an overall low probability mass, or are distributed unsystematically throughout the parameter space.

\section{Robustness to adversarial examples}

\begin{figure*}
  \includegraphics[width=0.49\textwidth]{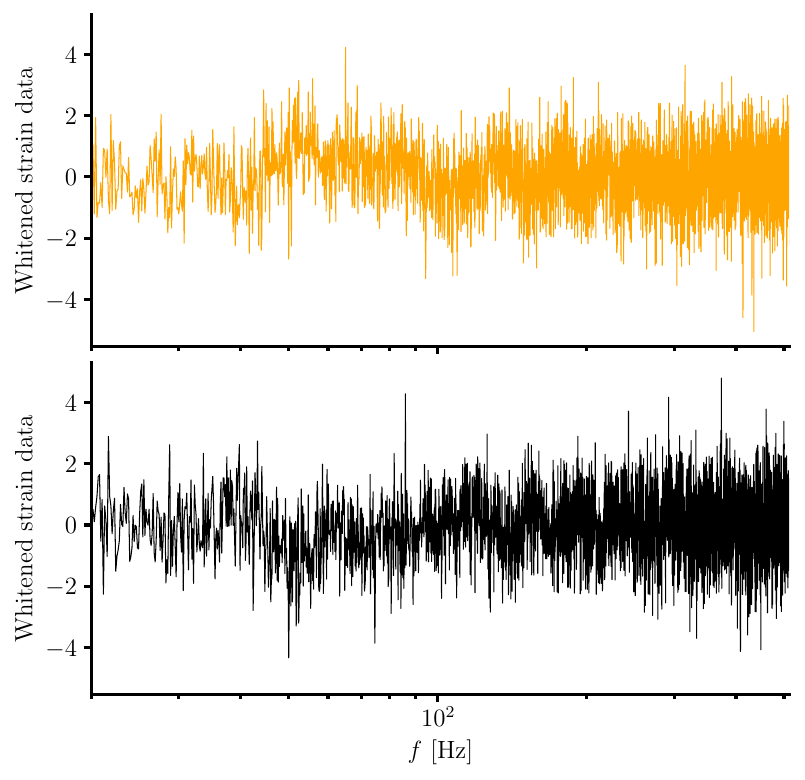}
  \hfill
  \includegraphics[width=0.49\textwidth]{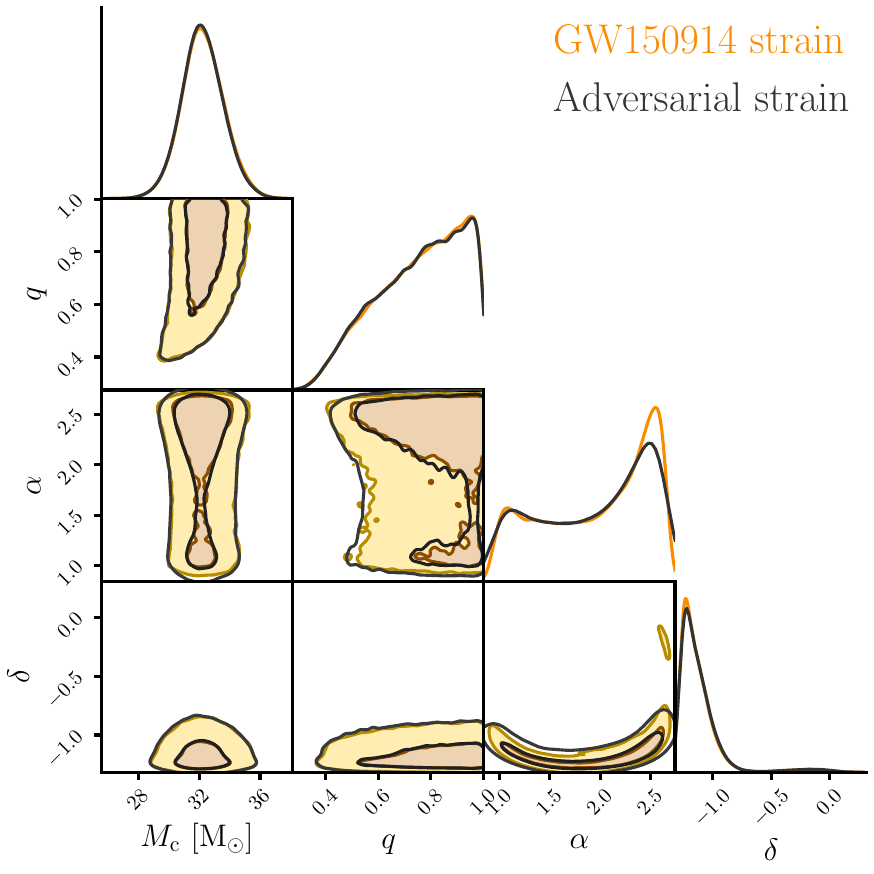}
  \caption{\label{fig:adverarial-example}Left: Strain data (real part) in the LIGO Hanford detector. The upper row shows the measured data for GW150914, the lower row shows an adversarial example that is synthetically generated to mislead the inference network. Right: The inference network infers almost identical posteriors for both strain datasets.
  }
\end{figure*}

An adversarial example~\cite{szegedy2013intriguing,goodfellow2014explaining} refers to data $d_\text{adv.}$ that is specifically designed to mislead a neural network. Such examples can be generated by following gradients of the network output (or some function thereof) starting from some real data $d_\text{true}$ and sequentially adding small perturbations to maximally change the output. Although the resulting adversarial example $d_\text{adv.}$ is often barely distinguishable from $d_\text{true}$, the neural network output can change dramatically.

In the context of posterior estimation, the output is a high-dimensional distribution which one can alter in multiple ways. We tried to shift or truncate the predicted \textsc{Dingo} distribution $q(\theta|d_\text{adv.})$ by applying only minimal modifications to the data. We found that \textsc{Dingo} is remarkably robust to such attacks, its output could barely be changed without significantly changing the input data $d$. This unusual robustness is attributed to two factors. First, the training data itself is very noisy, which regularizes \textsc{Dingo} models. Second, the first layer of \textsc{Dingo} networks is seeded with principal components of clean GW signals~\cite{Dax:2021tsq}, so adversarial perturbations are projected onto the manifold of GW signals.

We thus explore a slightly different notion of adversarial attacks. Starting from strain data $d$ initialized with random Gaussian noise, we aim to modify $d$ such that \textsc{Dingo} estimates identical posteriors for $d_\text{adv.}$ and the real strain data $d_\text{true}$ for GW150914. Specifically, we minimize the KL divergence $D_\text{KL}(q(\theta|d_\text{true})||q(\theta|d_\text{adv.}))$ via
\begin{equation}\label{eq:adversarial-example}
    d_\text{adv.} = \argmax_d \mathbb{E}_{\theta\sim q(\theta|d_\text{true})}\log q(\theta|d).
\end{equation}
In contrast to the technique mentioned above, we here do not constrain the difference between $d_\text{adv.}$ and $d_\text{true}$ to be small. To optimize~\eqref{eq:adversarial-example} we need to take gradients of the \textsc{Dingo} density with respect to $d$, which is intractable with the iterative GNPE~\cite{Dax:2021tsq,Dax:2021myb} method. Instead, we use a \textsc{Dingo} network trained with standard NPE. We use the adam~\cite{Kingma:2014vow} optimizer with a learning rate of 0.03 to optimize Eq.~\eqref{eq:adversarial-example} with 400 gradient steps (batch size 1024). The resulting strain $d_\text{adv}$ is visibly different from the true GW150914 strain $d_\text{true}$, but the estimated \textsc{Dingo} posteriors are almost identical (Fig.~\ref{fig:adverarial-example}). 

With \textsc{Dingo-IS}, we find a sample efficiency of $\epsilon=1.48\%$ for the real GW150914 strain $d_\text{true}$. This is substantially smaller than the sample efficiency achieved with GNPE ($\epsilon = 28.8\%)$, since standard NPE does not use the physical symmetries and is hence less accurate. However, the \textsc{Dingo-IS} posterior is still accurate and the evidence estimate ($\log p(d)=-15831.88\pm 0.03$) is in good agreement with the result reported in the main paper. For $d_\text{adv.}$ on the other hand, \textsc{Dingo-IS} achieves a sample efficiency of $\epsilon = 0.006\%$, clearly identifying the adversarial example as a \textsc{Dingo} failure case.

\section{Additional Results}

\begin{figure*}
  \includegraphics[width=\textwidth]{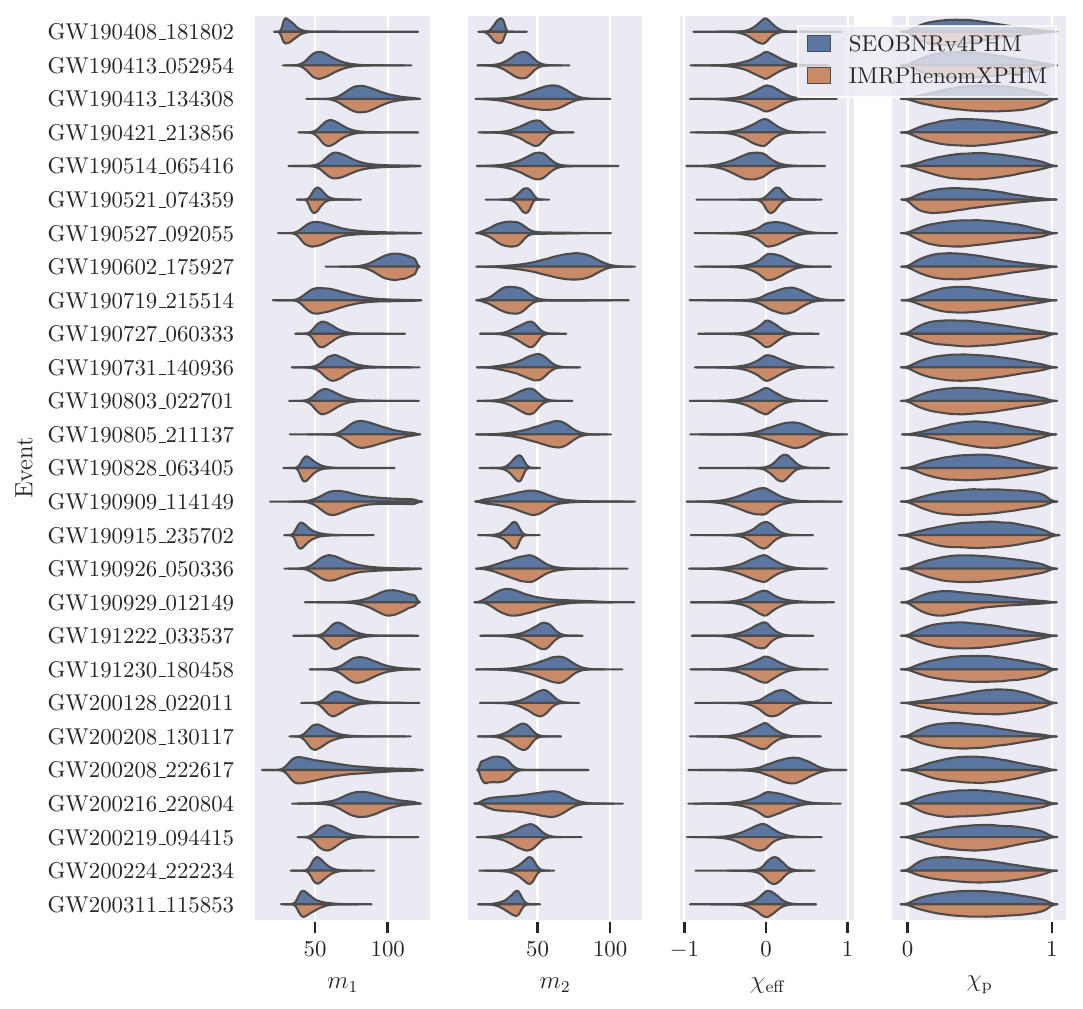}
  \caption{\label{fig:1d-marginals}Posterior distributions for
    component masses and effective spin parameters.
    We show all events from the main paper with $\epsilon > 2\%$ for both
    waveform models to ensure smooth posteriors.
    We observe good agreement between the two
    waveform models. In a future publication we will include a more complete
    catalog, which incorporates Virgo data, and includes a
    more careful treatment of noise artifacts and data conditioning.}
\end{figure*}

Fig.~\ref{fig:1d-marginals} shows one-dimensional marginal posteriors for a subset of GW
events analyzed in the main paper, comparing the two
waveform models IMRPhenomXPHM and SEOBNRv4PHM. We see that the models give results that
appear to be in good agreement.

\begin{figure*}
  \includegraphics[width=0.42\textwidth]{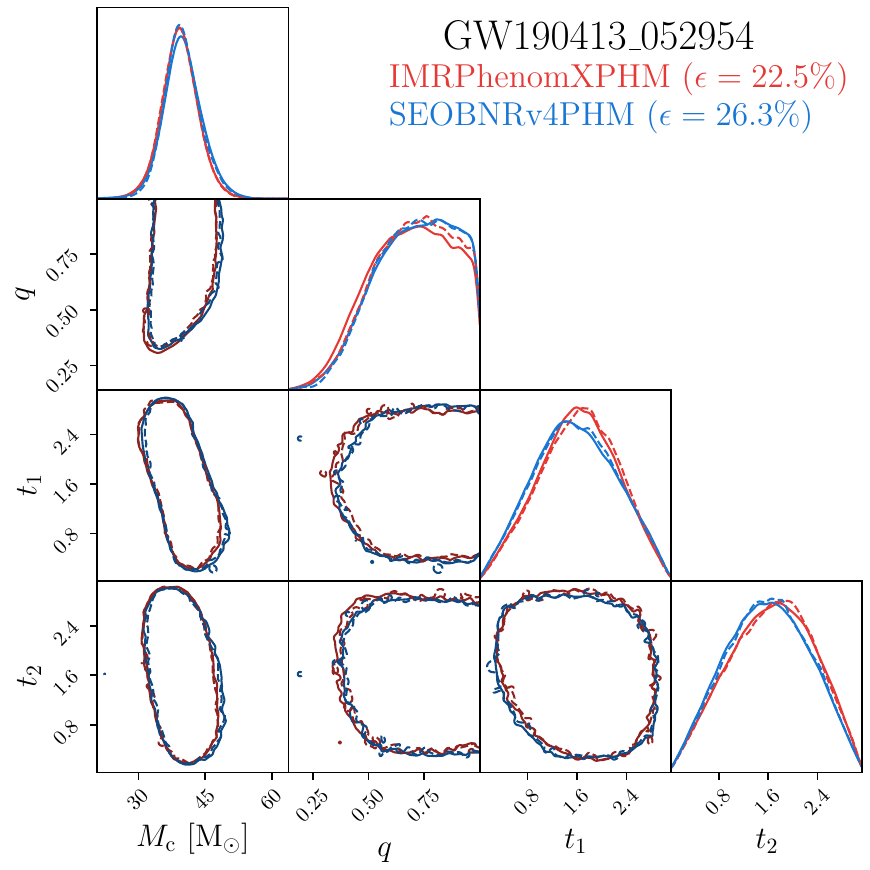}
  \hspace{1.3cm}
  \includegraphics[width=0.42\textwidth]{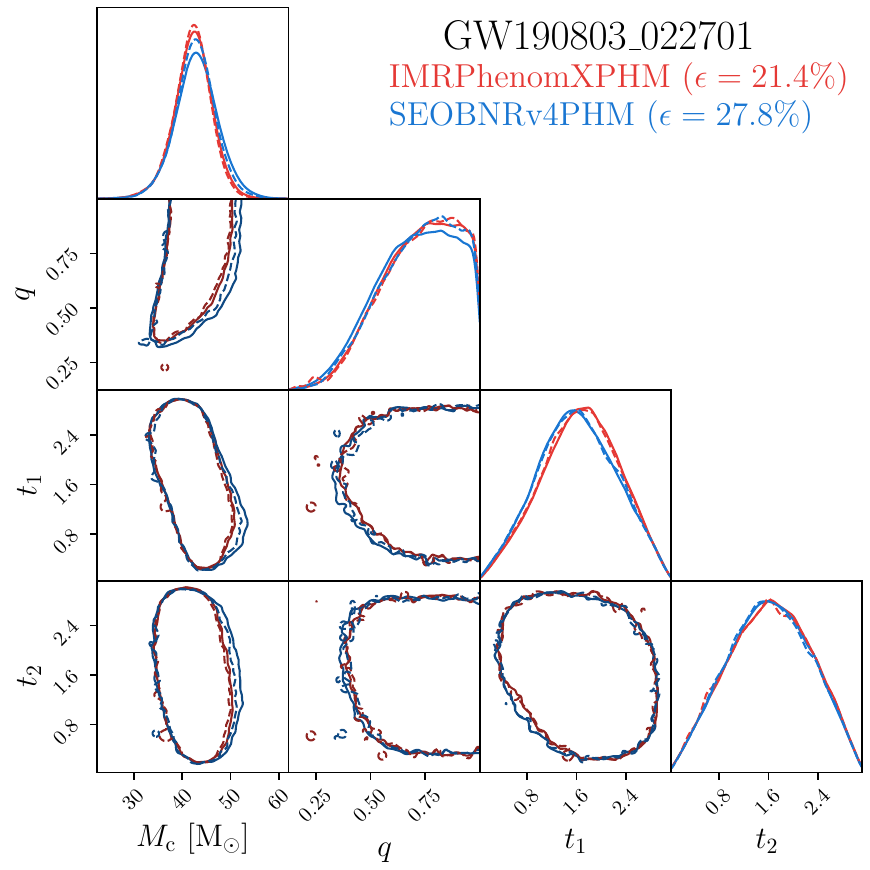}
  \includegraphics[width=0.42\textwidth]{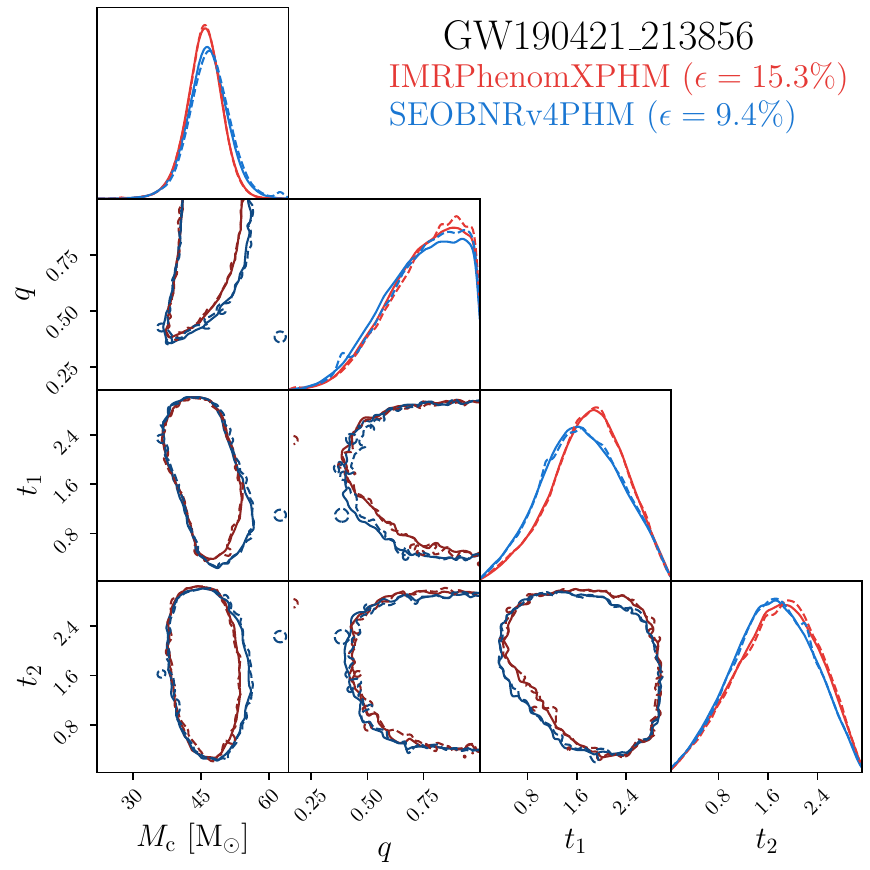}
  \hspace{1.3cm}
  \includegraphics[width=0.42\textwidth]{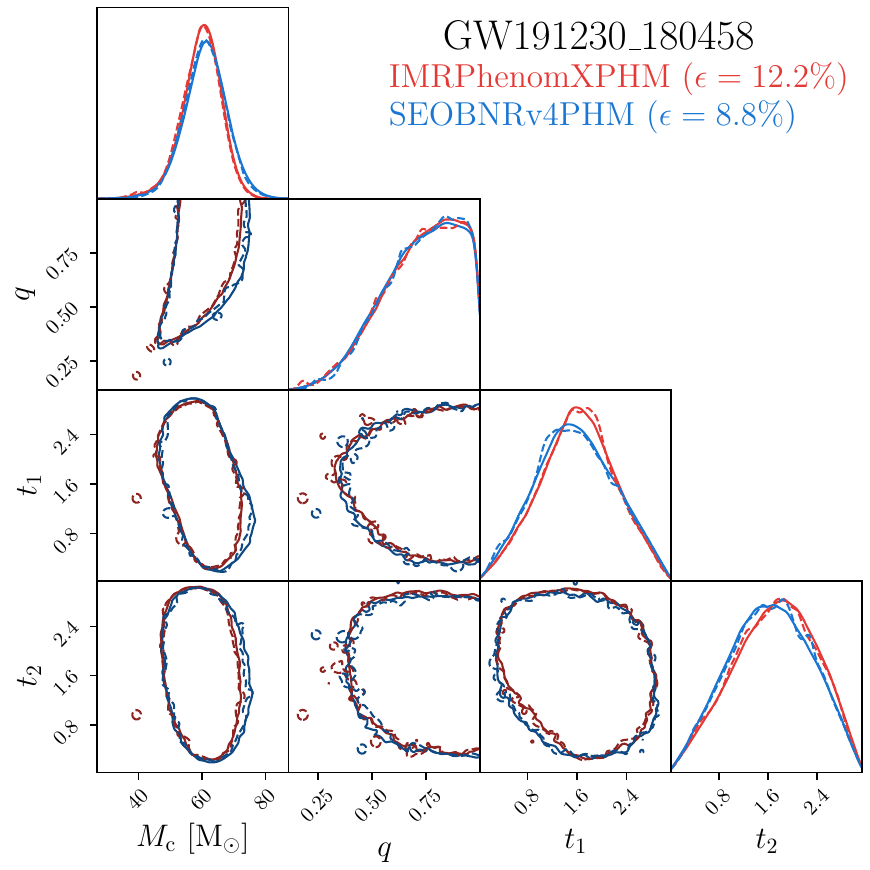}
  \includegraphics[width=0.42\textwidth]{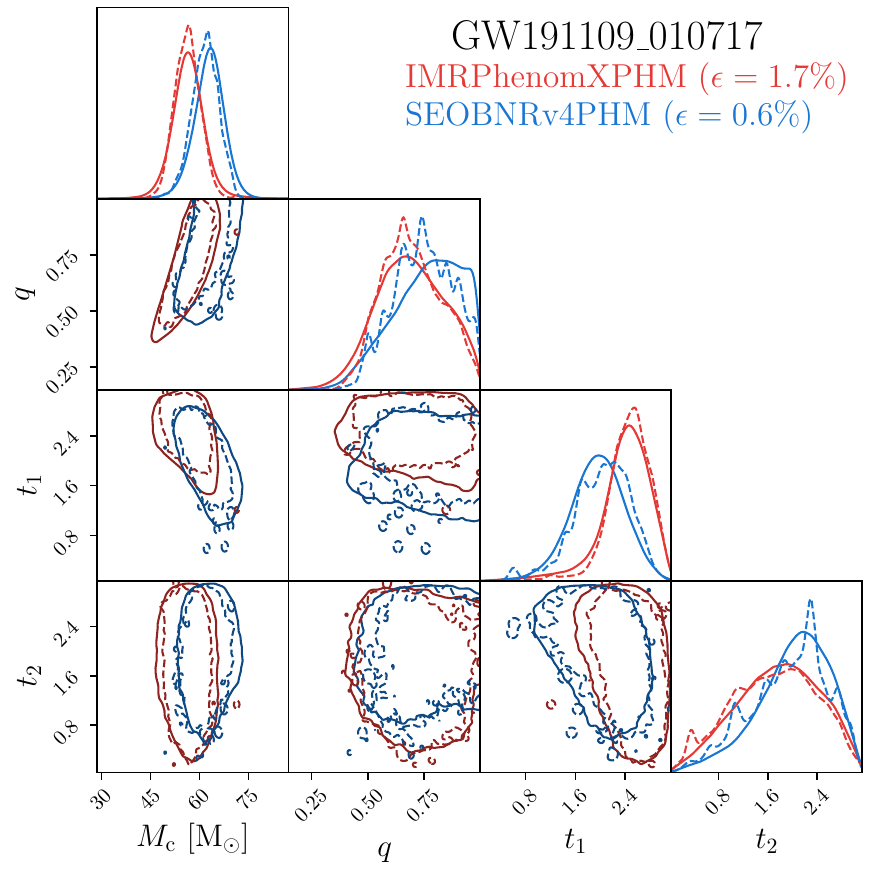}
  \hspace{1.3cm}
  \includegraphics[width=0.42\textwidth]{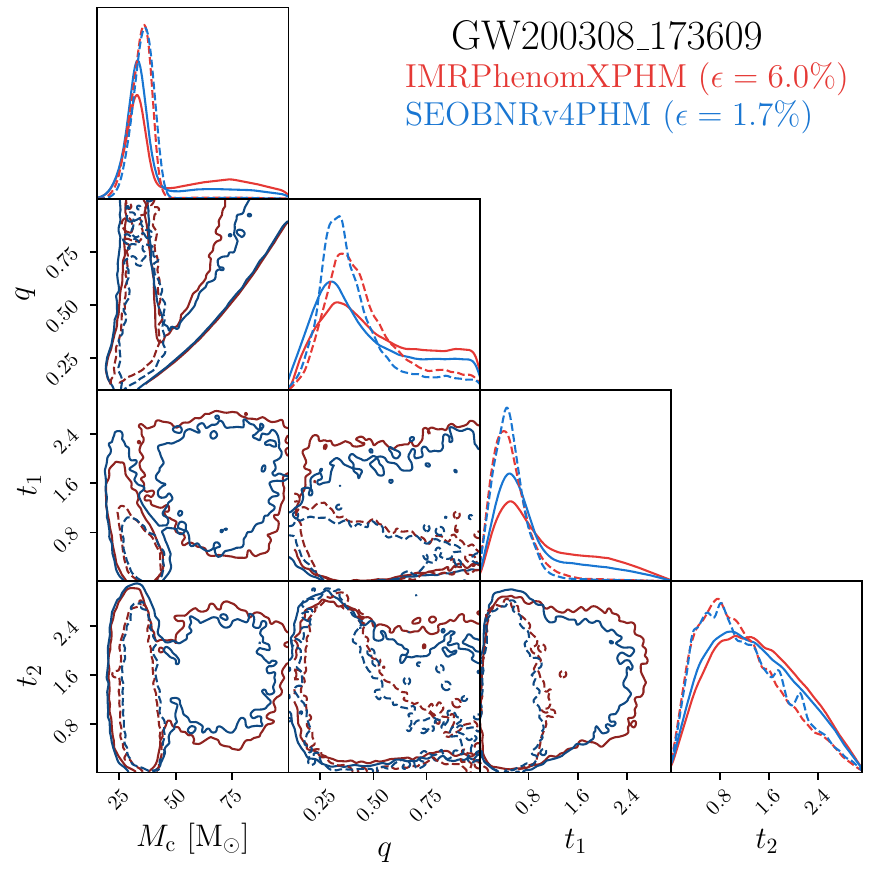}
  \caption{\label{fig:O3_cornerplots}Marginalized one- and two- dimensional posterior distributions for selected O3 events, comparing \textsc{Dingo} (solid lines) and \textsc{Dingo-IS} (dashed) inference results with waveform models IMRPhenomXPHM and SEOBNRv4PHM. Contours represent 90\% credible regions. For events with high (upper row) or medium (middle row) sample efficiency, the initial \textsc{Dingo} results are often accurate and only deviate slightly from \textsc{Dingo-IS} results. For events with low effective sample size (lower row), the \textsc{Dingo-IS} contours are often not smooth. Yet, the initial \textsc{Dingo} results may capture the marginals well, see GW191109\_010717.
  }
\end{figure*}

Fig.~\ref{fig:O3_cornerplots} shows posterior marginals for several GW events from O3. A large sample efficiency often corresponds to good agreement of the \textsc{Dingo} and \textsc{Dingo-IS} marginals. The sample efficiency is sensitive to deviations in the full 15 dimensional parameter space, so small sample efficiencies do not necessarily imply inaccurate \emph{marginal} distributions.

\end{document}